\documentclass[journal]{IEEEtran}
\usepackage{cite}
\usepackage{amsmath} 
\usepackage{subfigure} 
\usepackage{graphicx}
\usepackage{enumerate}
\usepackage{enumitem}%resume the number
\usepackage{bm}
\usepackage{amsfonts}
\usepackage{booktabs}
\usepackage[ruled]{algorithm2e}

\begin{document}
\title{Computation Rate Maximum for Mobile Terminals in UAV-assisted Wireless Powered MEC Networks with Fairness Constraint}

\author{
        Xiaoyi~Zhou,
        Liang~Huang,~\IEEEmembership{Member,~IEEE,}
        Tong~Ye,~\IEEEmembership{Member,~IEEE,}
        Weiqiang~Sun,~\IEEEmembership{Senior~Member,~IEEE}% <-this % stops a space
\thanks{Xiaoyi Zhou, Tong Ye, and Weiqiang Sun are with the State Key Laboratory of Advanced Optical Communication Systems and Networks, Shanghai Jiao Tong University, Shanghai 200240, China (e-mail: zhouxiaoyi@sjtu.edu.cn; yetong@sjtu.edu.cn; sunwq@sjtu.edu.cn).}%
\thanks{Liang Huang is with the College of Information Engineering, Zhejiang University of Technology, Hangzhou, China 310058 (e-mail: lianghuang@zjut.edu.cn).}}
% \thanks{Manuscript received April 19, 2005; revised August 26, 2015.}}

% \markboth{Journal of \LaTeX\ Class Files,~Vol.~14, No.~8, August~2015}%
% {Shell \MakeLowercase{\textit{et al.}}: Bare Demo of IEEEtran.cls for IEEE Journals}

\maketitle

\begin{abstract}
This paper investigates an unmanned aerial vehicle (UAV)-assisted wireless powered mobile-edge computing (MEC) system, where the UAV powers the mobile terminals by wireless power transfer (WPT) and provides computation service for them. We aim to maximize the computation rate of terminals while ensuring fairness among them. Considering the random trajectories of mobile terminals, we propose a soft actor-critic (SAC)-based UAV trajectory planning and resource allocation (SAC-TR) algorithm, which combines off-policy and maximum entropy reinforcement learning to promote the convergence of the algorithm. We design the reward as a heterogeneous function of computation rate, fairness, and reaching of destination. Simulation results show that SAC-TR can quickly adapt to varying network environments and outperform representative benchmarks in a variety of situations.
\end{abstract}


\IEEEpeerreviewmaketitle

\section{Introduction}
\IEEEPARstart{I}{n} recent years, unmanned aerial vehicles (UAV)-assisted wireless powered mobile-edge computing (MEC) network has attracted more and more attention \cite{5,6,7,8,9}. Due to technological advances, today’s UAVs can equip MEC servers with strong computing capabilities and energy transmitters. It can perform not only computation offloading via MEC technology \cite{32} but also wireless charging via wireless power transfer (WPT) \cite{33} for mobile terminals, which need to operate computation-intensive applications but have limited computing capacity and battery lifetime. The UAV is thus suitable for building temporary MEC systems for mobile terminals in some special situations. For example, the UAV can provide service for the mobile terminals in the scenarios where base station (BS) is damaged, in the public meeting places where there is a traffic hotspot, or in the remote fields where there is a coverage hole of wireless networks.

The UAV-assisted wireless powered MEC networks were previously investigated for ground terminals with fixed locations \cite{5,6,7,8,9}. To maximize the network utility in terms of computation rate \cite{5,9} or minimize the energy consumption \cite{6,7,8}, previous works optimized UAV trajectory, offloading decision, and resource allocation. Sometimes, the UAV is required to arrive at specified locations automatically, so that it can be utilized in the places that are difficult for people to reach, thereby reducing labor costs \cite{5,8}. The previous works solved the optimization problems by offline algorithms such as successive convex approximation \cite{5,6,8} and block coordinate descending \cite{9}. The numerical results in \cite{5,6,7,8,9} show that these algorithms work well in the scenarios where the locations of terminals are fixed. 

However, terminals in practice such as smartphones, tablets, wearable devices, and tracking collars carried by wildlife \cite{2} are typically in motion, and their trajectories are likely to be stochastic. To serve mobile terminals, online algorithms are needed to make decisions based on real-time information. Unfortunately, the existing algorithms designed for the terminals with fixed locations \cite{5,6,7,8,9} are offline algorithms, and may not work well in these scenarios since all of them need environment information a priori \cite{31}.

In this paper, we study a UAV-assisted wireless powered MEC network where a flying UAV serves multiple mobile terminals. We aim to maximize the computation rate of all terminals while ensuring the fairness among them. Herein, considering fairness is to balance the computation performance of different terminals. We demonstrate that this problem is a joint optimization and continuous control problem of the UAV trajectory and the resource allocation of terminals, under the condition that the trajectories of terminals are stochastic. Therefore, we propose a soft actor-critic (SAC) based deep-reinforcement-learning (DRL) algorithm for trajectory planning and resource allocation (SAC-TR). Since this problem is a complex high-dimensional DRL task, SAC-TR combines off-policy and maximum entropy reinforcement learning to ensure sampling efficiency and stabilize convergence at the same time. Taking the computation rate, fairness, and reaching destination into consideration, we design the reward in SAC-TR as a heterogeneous function to satisfy multiple objectives simultaneously. The simulation results show that SAC-TR outperforms representative benchmarks in most cases.

The main contributions of this paper are highlighted as follows.
\begin{enumerate}[]
\item To the best of our knowledge, we are the first to provide an online algorithm for trajectory planning and resource allocation for mobile terminals in the UAV-assisted wireless powered MEC network.
\item By integrating a fairness index into the reward of SAC-TR, we guarantee the fairness among different terminals according to the needs of scenarios.
\item By using the progress estimate in the reward of SAC-TR, the UAV can reach the specified destination automatically and the convergence of SAC-TR is accelerated.
\item SAC-TR can converge steadily and fast adapt to unexpected changes of the environment.
\end{enumerate}

The rest of this paper is organized as follows. In Section \ref{related}, we review the related works. We model the UAV-assisted wireless powered MEC network in Section \ref{network} and formulate the trajectory planning and resource allocation problem in Section \ref{formulation}. In Section \ref{algorithm}, the detailed design of SAC-TR is given. We evaluate the performance of our algorithm by simulations in Section \ref{evaluation}. Finally, the paper is concluded in Section \ref{conclusion}.

\section{Related Works}\label{related}
The previous works related to our paper include those focusing on wireless powered MEC networks \cite{13,14,16,19,21,5,6,7,8,9}, and UAV-assisted communication networks for mobile terminals \cite{3,22,4}.

\subsection{Wireless Powered MEC Network}\label{related-mec}
The wireless powered MEC network can be divided into two types according to the carriers of MEC servers and energy transmitters. The first type is the wireless powered MEC network, where the BS is the carrier \cite{13,14,16,19,21}, while the second type is the UAV-assisted wireless powered MEC network, where the UAV is the carrier \cite{5,6,7,8,9}.

In the system with the BS as the carrier, the works mainly focus on resource allocation and offloading decision \cite{13,14,16,19,21}. In \cite{13} and \cite{14}, the parameters such as offloading decision, CPU frequency, and transmission time of terminals were optimized to minimize the energy consumption and maximize the computation rate respectively. Different from \cite{13,14}, Mao et al. investigated the max-min energy efficiency optimization problem to guarantee the fairness of energy efficiency among different devices \cite{16}. To utilize the advantages of DRL in handling problems with sophisticated state space and time-varying environment, Min et al. \cite{19} proposed a deep Q network (DQN) based offloading policy for energy-harvesting MEC network to improve the computation performance. Huang et al. \cite{21} proposed a DRL-based online computation offloading (DROO) framework. Instead of solving for the hybrid integer continuous solution altogether, DROO decomposes the optimization problem into a binary offloading decision sub-problem and a continuous resource allocation subproblem, and tackles them separately by deep learning and traditional optimization methods, respectively. 

In the system with the UAV as the carrier, the previous works only considered the case where the locations of terminals are fixed \cite{5,6,7,8,9}. To maximize the weighted sum computation rate of terminals, Zhou et al. \cite{5} jointly optimized the CPU frequencies, transmit powers, and offloading times of terminals as well as the UAV trajectory. Ref. \cite{6} minimized the energy consumption of the UAV while guaranteeing the computation rate of all terminals. Ref. \cite{7} proposed a time-division multiple access (TDMA) based workflow model, which allows parallel transmitting and computing. In particular, the UAV was arranged to hover over designated successive positions, and the parameters such as the service sequence of terminals, computing resource allocation, and hovering time of UAV were jointly optimized. To assist the service of UAV, Liu et al. \cite{8} utilized idle sensor devices to cooperate with the UAV to provide computation offloading service for busy sensor devices, and Hu et al. \cite{9} utilized access points (APs) to offer wireless power and computation offloading services for the UAV. The offline algorithms in \cite{5,6,7,8,9} require the system information a priori.

\subsection{UAV-assisted Communication Network for Mobile Terminals}\label{related-com}
The UAV-assisted communication network for mobile terminals is similar to the UAV-assisted MEC network for mobile terminals. The difference is that the UAV in the former case carries out traffic offloading while that in the latter case performs computation offloading. The major problem in the UAV-assisted communication network is resource allocation and UAV trajectory design for performance optimization. Ref.\cite{3} proposed a deterministic policy gradient (DPG) based algorithm to maximize the expected uplink sum rate of terminals. Ref.\cite{22} considered the scenario where a group of UAVs are employed to enhance communication coverage area. This paper proposed an actor-critic (AC) based algorithm to optimize the UAV trajectory, such that the objectives including coverage expansion, fairness improvement, and power saving can be achieved. However, DPG \cite{3} and AC \cite{22} are hard to converge if applied to the complex high-dimensional DRL task, e.g., the problem considered in this paper, which jointly optimizes multiple types of parameters. Ref.\cite{4} aimed to maximize the throughput of a UAV-assisted cellular offloading network. Ref.\cite{4} discretized the flight direction of UAV and the transmit power of terminals and devised a value-based DRL algorithm. Since this algorithm has to search the action space exhaustively in each iteration, it cannot be used for problems with high-dimensional or continuous actions \cite{20}.

\section{UAV-assisted Wireless Powered MEC Network}\label{network}
\begin{figure}[!t]
\centering
\includegraphics[width=3in]{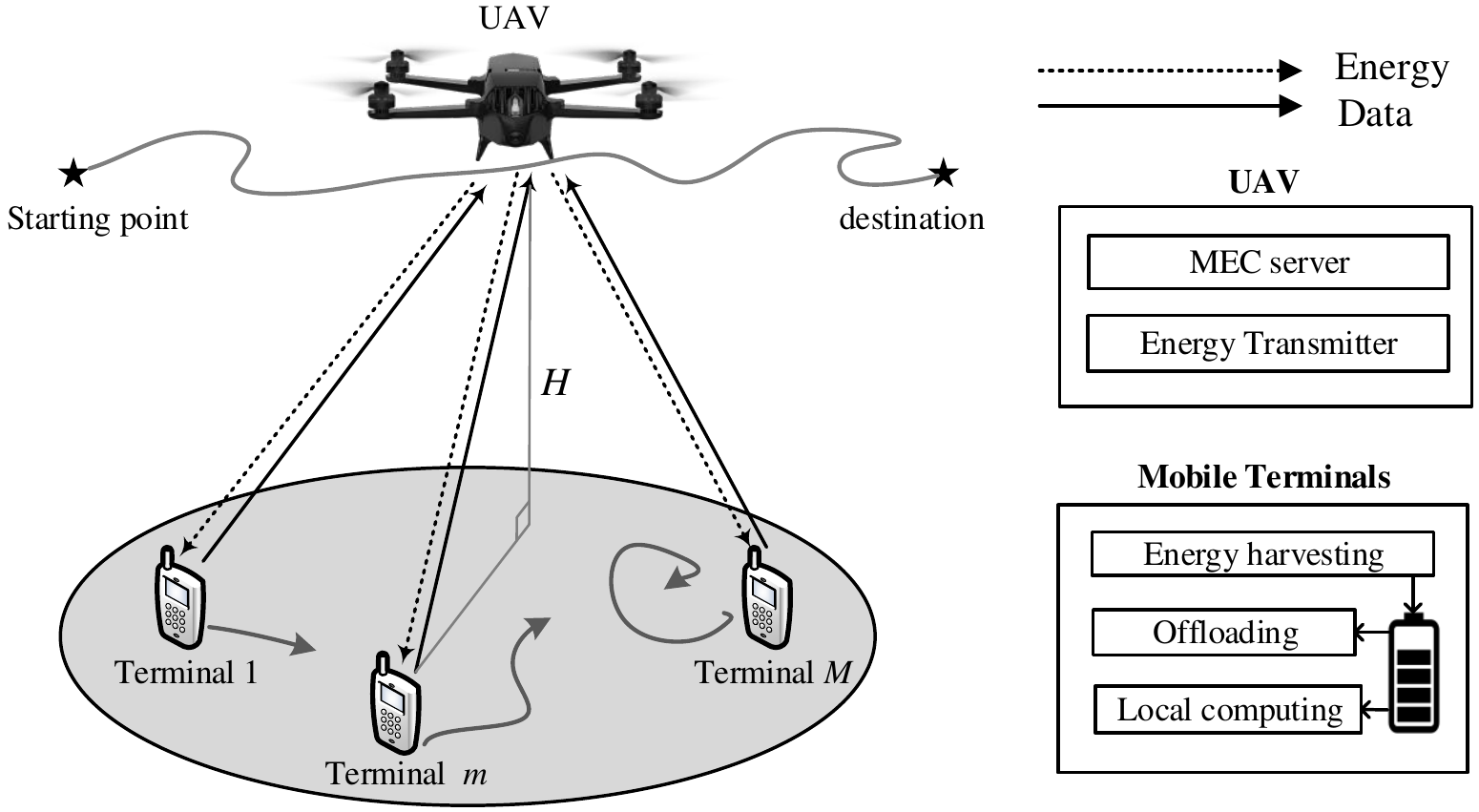}
\caption{The UAV-assisted wireless powered MEC network with mobile terminals.}
\label{fig1}
\end{figure}

Fig.\ref{fig1} illustrates the UAV-assisted wireless powered MEC network considered in this paper.There are a set of mobile terminals, denoted by $\mathcal{M}=\left\{1,2,\cdots,M\right\}$ and a UAV. All the terminals move on the ground with altitude 0, and the UAV flies at a fixed altitude, denoted by $H$, such that it can avoid frequent ascent and descent to evade surficial obstacles. The UAV is equipped with MEC servers and energy transmitters and serves these mobile terminals with low battery lifetime and computing capacity. Each terminal has accumulated computation tasks, which can be divided into two parts. One part is executed locally by the mobile terminal and the other part is offloaded to and executed at the UAV, which is known as the partial offloading mode. In the meanwhile, the UAV broadcasts radio-frequency (RF) energy to all mobile terminals, and terminals harvest the energy and store it in the chargeable battery. The UAV/terminal can perform energy transferring/harvesting, computing, and data exchange simultaneously \cite{13,14,16,5}. The UAV is required to arrive at a designated location at the end of the flight \cite{5,8}.

\subsection{Computation Offloading}\label{network-offloading}
\begin{figure}[!t]
\centering
\includegraphics[width=3in]{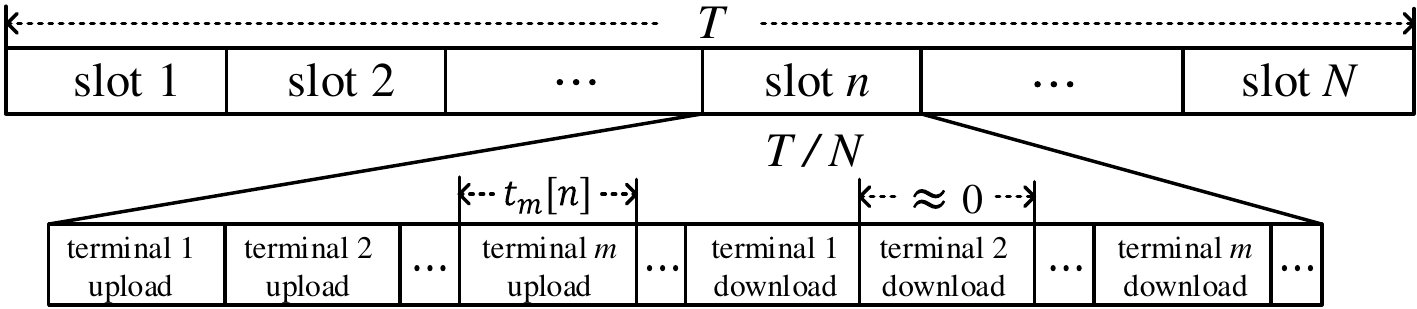}
\caption{Mobile terminals adopt the TDMA protocol to communicate with the UAV.}
\label{fig2}
\end{figure}

Mobile terminals adopt the TDMA protocol to communicate with the UAV, as illustrated in Fig.\ref{fig2}. The flight time of the UAV, denoted by $T$, is discretized into $N$ slots. The duration of a time slot is very short such that the locations of UAV and terminals and the channel gain almost keep unchanged. In each slot, the mobile terminals offload computation tasks to the UAV in a round-robin manner and download the computation results from it after completion.

\subsubsection{Computation Time and Data-exchange Time}
We denote $t_m[n]$ as the proportion of uploading time of the $m$th terminal in slot $n$. In general, the computing capacity of the MEC servers on the UAV is powerful and the size of computation result is quite small. Thus, we assume that
\begin{enumerate}[label=(a\arabic*)]
\item the computation time of UAV and the result downloading time of terminals can be neglected \cite{5,14,16}.\label{a1}
\end{enumerate}

\subsubsection{Channel Condition}
The data exchange between the UAV and the terminals is influenced by the wireless channel conditions. In our model, we assume that
\begin{enumerate}[resume,label=(a\arabic*)]
\item the channels is depicted by the air-to-ground model \cite{10,35},\label{a2}
\item the impact of the Doppler effect in data exchange due to the position changes of the UAV and mobile terminals can be perfectly compensated by the receivers, and\label{a3}
\item the path loss is dominated by the large-scale fading of channels between the UAV and terminals \cite{5,35}.\label{a4}
\end{enumerate}

Employing the three-dimensional (3D) Euclidean coordinate, we let $\mathbf q_m[n]\!=\!\left(x_m[n],y_m[n]\right)$ be the horizontal plane coordinate of the $m$th terminal and $\mathbf q_u[n]\!=\!\left(x_u[n],y_u[n]\right)$ be that of the UAV in slot $n$. Under the air-to-ground model, the path loss between the $m$th terminal and the UAV in the $n$th slot is given by \cite{10}, as
\begin{equation}\label{eq1}
\begin{split}
L_m[n]=20lg\left(\frac{4\pi f_c\left({\Vert \mathbf q_u[n]-\mathbf q_m[n]\Vert}^2+H^2\right)^{\frac{1}{2}}}{c}\right)&\\
+P_{LoS}\eta_{LoS}+\left(1-P_{LoS}\right)\eta_{NLoS},&
\end{split}
\end{equation}
where $\Vert \cdot \Vert$ is Euclidean norm, $f_c$ is the carrier frequency, $c$ is the light speed, $P_{LoS}$ is the probability that the link between the terminal and the UAV is a Line-of-Sight (LoS) link, and $\eta_{LoS}$ and $\eta_{NLoS}$ are the additional loss caused by the LoS and non-LOS (NLoS) link on the top of the free space path loss. The values of $\eta_{LoS}$ and $\eta_{NLoS}$ are determined by the environments, such as urban and rural. According to \cite{10}, $P_{LoS}$ is given by
\begin{equation}\label{eq2}
P_{LoS}\!=\!\frac{1}{1\!+\!h\text{exp}\left(\!-l\!\left(\frac{180}{\pi}\text{arctan}\left(\frac{H}{\Vert\mathbf q_u[n]-\mathbf q_m[n] \Vert}\right)\!-\!h\!\right)\right)},
\end{equation}
where $h$ and $l$ are two constants determined by the environments \cite{10}. Accordingly, the channel power gain between the $m$th terminal and the UAV in slot $n$ is given by
\begin{equation}\label{eq3}
G_m[n]=10^{-L_m[n]/10}.
\end{equation}
\subsubsection{Task Offloading by Terminals}
The computation tasks that the terminal uploads to the UAV including the raw data and the communication overhead such as the encryption and the packet header \cite{14}. We assume that 
\begin{enumerate}[resume,label=(a\arabic*)]
\item each bit of raw data needs $\delta$ bits of upload data.
\end{enumerate}
Recall that the $m$th terminal offloads data to the UAV with duration $t_m[n]\cdot T/N$ in slot $n$. The volume of raw data that terminal $m$ offloads to the UAV in slot $n$ is
\begin{equation}\label{eq4}
U_m^o[n]=\frac{T}{N\delta}t_m[n]B\textrm{log}_2\left(1+\frac{P_m[n]G_m[n]}{\sigma^2}\right),
\end{equation}
where $P_m[n]$ is the transmit power of terminal $m$ in slot $n$, $B$ is the offloading bandwidth, and $\sigma^2$ is the noise power at the terminal. It follows that the energy consumption for offloading these data is $\frac{T}{N}t_m[n]P_m[n]$.

\subsection{Local Computation} \label{network-local}
Mobile terminals execute local computation tasks and adjust the CPU frequency by dynamic voltage and frequency scaling technique in each slot \cite{5,14,16}. Let $f_m[n]$ be the CPU frequency (unit: cycle/s) of the $m$th terminal in slot $n$, $C$ be the number of CPU cycles required for computing one bit of raw data. Then, the local computation bits of the $m$th terminal in the $n$th slot is
\begin{equation}\label{eq5}
U_m^l[n]=\frac{Tf_m[n]}{NC},
\end{equation}
Accordingly, the local energy consumption of the $m$th terminal in the $n$th slot is given by $\frac{T}{N}\zeta_cf_m^3[n]$, where $\zeta_c$ is the effective capacitance coefficient of the processor chip \cite{14}.

Let $U_m[n]$ be the number of computation bits of the $m$th terminal in the $n$th slot, including both the local and the offloaded ones. $U_m[n]$ is given by
\begin{equation}\label{eq6}
\begin{split}
U_m[n]=&U_m^l[n]+U_m^o[n]\\
=&\frac{Tf_m[n]}{NC}\!+\!\frac{T}{N\delta}t_m[n]B\textrm{log}_2\left(\!1\!+\!\frac{P_m[n]G_m[n]}{\sigma^2}\!\right)\!.
\end{split}
\end{equation}
Thus, the total computation bits of the $m$th terminal in the entire flight time are $U_m=\sum_{n=1}^{N}U_m[n]$.

\subsection{Wireless Power Transfer}\label{network-power}
The UAV broadcasts RF energy to all mobile terminals continuously during its flight time. We assume that
\begin{enumerate}[resume,label=(a\arabic*)]
\item the energy of the UAV is sufficient, and 
\item the transmit power of the UAV is a constant, $P_e$.
\end{enumerate}
The energy harvested by the $m$th terminal in slot $n$ is $\eta_0\frac{T}{N}G_m[n]P_e$, where $0\!<\!\eta_0\!\le\!1$ is the energy conservation efficiency.

\section{Problem Formulation}\label{formulation}
To ensure the performance of each terminal, we aim to maximize the sum computation bits of all mobile terminals in the entire flight time $\sum_{m=1}^{M}U_m$, while guaranteeing the fairness of computation bits among different terminals. Based on the Jain’s fairness index \cite{38}, we define the fairness index $I$ ($1/M\le I\le1$) as
\begin{equation}\label{eq7}
I=\frac{\left(\sum_{m=1}^{M}U_m\right)^2}{M\cdot\sum_{m=1}^{M}U_m^2}.
\end{equation}
Clearly, a larger $I$ indicates higher fairness. Accordingly, we define the objective function as a joint function of the computation bits and fairness, as $I^\omega\!\cdot\!\sum_{m=1}^{M}U_m$ , where $\omega$ is a non-negative integer used to adjust the proportion of $I$ in the objective function.

We intend to optimize the UAV trajectory and the resource allocation of terminals during the flight time of the UAV. Let $v_u[n]$ and $\theta_u[n]$ be the flight speed and direction of the UAV in slot $n$, respectively. The UAV trajectory is described by $\mathbf v_u=\left\{v_u[n]|n\in\mathcal{N}\right\}$ and $\bm\theta_u=\left\{\theta_u[n]|n\in\mathcal{N}\right\}$, where $\mathcal{N}=\left\{1,2,\cdots,N\right\}$. The resource allocation variables include the transmit powers, offloading times, and CPU frequencies in all the $N$ slots. In particular, the resource allocation variables are $\mathbf P=\left\{P_m[n]|m\in\mathcal{M},n\in\mathcal{N}\right\}$, $\mathbf t=\left\{t_m[n]|m\in\mathcal{M},n\in\mathcal{N}\right\}$, and $\mathbf f=\left\{f_m[n]|m\in\mathcal{M},n\in\mathcal{N}\right\}$.
Note that, the transmit power and offloading time affect the offloading performance, and the CPU frequency decides the number of local computation bits. Consequently, to maximize the objective function, we should jointly optimize the flight speed and  direction of the UAV, and the transmit powers, offloading times, and CPU frequencies of mobile terminals in each slot.

Our optimization problem is formulated as
\begin{subequations}\label{eq8}
\begin{align}
\mathbf P_1: &\max_{\mathbf v_u,\bm\theta_u,\mathbf p,\mathbf t,\mathbf f}I^\omega\cdot\sum_{m=1}^{M}\sum_{n=1}^{N}U_m[n]\\
\textrm{s.t.}\ &C1:P_m[n]\ge0,\ f_m[n]\ge0,\quad m\in\mathcal{M},n\in\mathcal{N},\\
\begin{split}
&C2: \sum_{i=1}^{n}\frac{T}{N}\left(\zeta_cf_m^3[n]+t_m[n]P_m[n]\right)\le e_m\\
&\quad\quad+\sum_{i=1}^{n}\eta_0\frac{T}{N}G_m[n]P_e,\quad m\in\mathcal{M},n\in\mathcal{N},
\end{split}\\
&C3: \sum_{m=1}^{M}t_m[n]\le1,\quad n\in\mathcal{N},\\
&C4: 0\le v_u[n]\le v_u^{\textrm{max}},\quad n\in\mathcal{N},\\
&C5: 0\le \theta_u[n]\le 2\pi,\quad n\in\mathcal{N},\\
&C6: \mathbf q_u[1]=\mathbf q_S,\\
&C7: \mathbf q_u[N+1]=\mathbf q_D,
\end{align}
\end{subequations}
where $e_m$ is the initial energy of the $m$th terminal, $v_u^{max}$ is the maximum horizontal flying speed of UAV, $\mathbf q_u[1]$ and $\mathbf q_u[N+1]$ are respectively the locations of UAV in the first slot and after the last slot, $\mathbf q_S$ and $\mathbf q_D$ are the locations of designed starting point and the destination.

$C1$ indicates that the transmit powers and CPU frequencies of terminals should be non-negative. $C2$ restricts that, by each slot, the accumulated energy consumption of a terminal cannot exceed the sum of the initial energy and the energy harvested by this terminal. $C3$ states that the sum of offloading time of all terminals in each slot cannot exceed the duration of a slot. $C4$ and $C5$ give the range of the flight speed and direction of the UAV. $C6$ and $C7$ restrict the starting point and the destination of the UAV.

Though problem $\mathbf P_1$ is a sequential decision problem that can be characterized by a Markov decision process (MDP), the moving trajectories of terminals may be unpredictable and cannot be known in advance. Also, it involves the joint optimization and continuous control of high-dimensional parameters. As a result, traditional optimization approaches fail to solve this problem. For example, offline algorithms such as dynamic planning, successive convex approximation, or block coordinate descending, require the system information a priori; the DQN method \cite{17} can only deal with problems with discrete or low-dimensional actions \cite{20}; also, it is a challenge for the policy gradient method \cite{26} or the actor-critic (AC) method \cite{25} to maintain both high sample efficiency and stable convergence at the same time \cite{30}, when they are employed to handle a $\mathbf P_1$-like complex high-dimensional DRL task. We thus will introduce the soft actor-critic (SAC) method \cite{1} to solve this problem in the next section.

\section{SAC-based Algorithm for Trajectory Planning and Resource Allocation}\label{algorithm}
In this section, we propose an SAC-based trajectory planning and resource allocation (SAC-TR) algorithm to solve problem $\mathbf P_1$. To deal this complex high-dimensional DRL task, SAC-TR adopts the combination of off-policy and maximum entropy reinforcement learning in SAC method, so as to increase sampling efficiency and stabilize convergence at the same time. Taking into consideration the computation rate, fairness, and reaching of destination, we design a heterogeneous reward function in SAC-TR. SAC-TR is introduced in the following three parts. We present the main design of SAC-TR in Section \ref{algorithm-design} and the heterogeneous reward function in Section \ref{algorithm-reward}. Section \ref{algorithm-entropy} introduces the maximum entropy reinforcement learning and gives the gradient descent formulas of the neural networks in SAC-TR.

\subsection{Design of SAC-TR}\label{algorithm-design}
\begin{figure}[!t]
\centering
\includegraphics[width=3.3in]{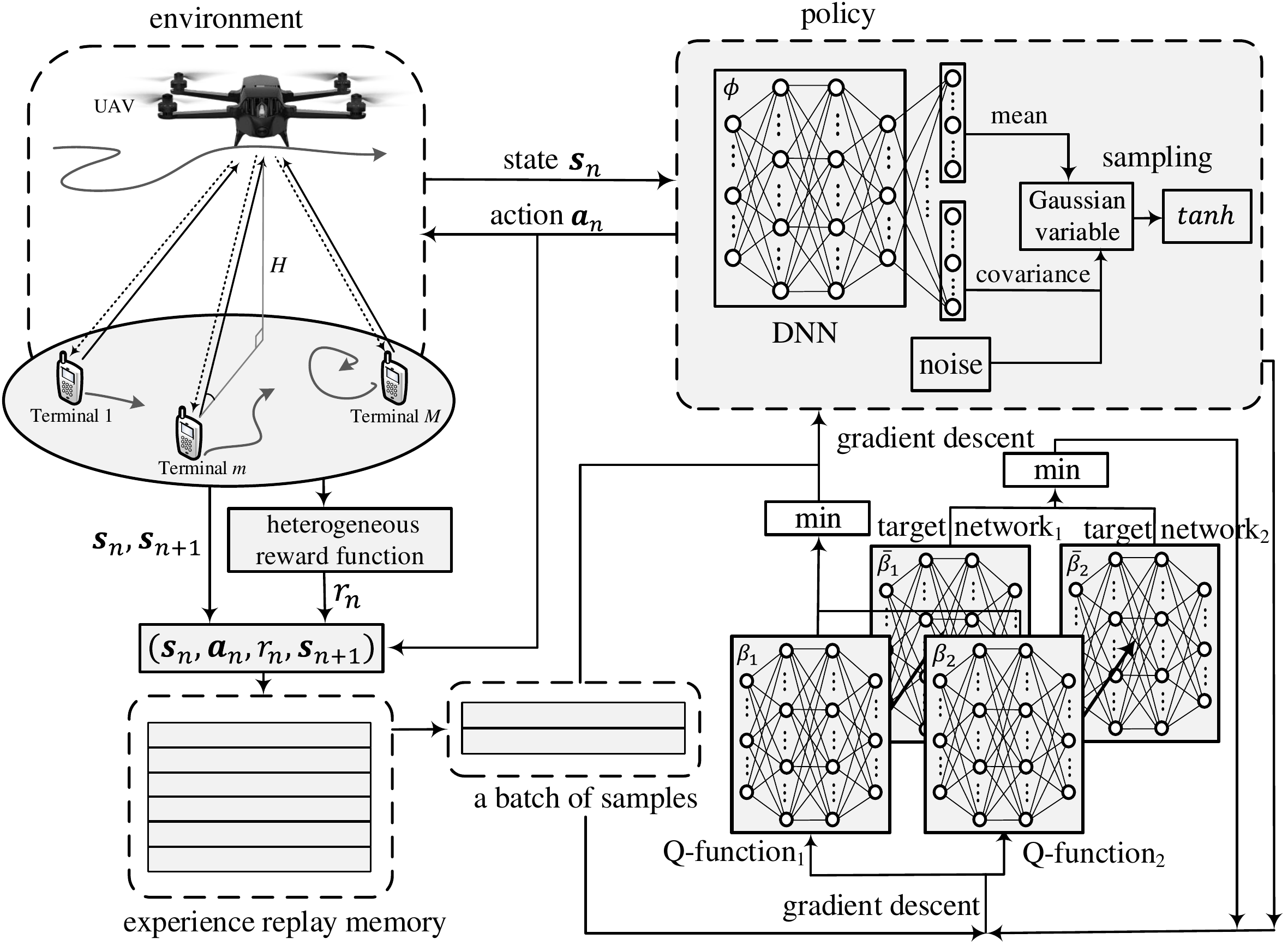}
\caption{Sketch of SAC-TR.}
\label{fig3}
\end{figure}

Fig.\ref{fig3} plots the structure of SAC-TR, which consists of a policy function, denoted by $\pi_\phi\!\left(\!\bm a_n|\bm s_n\!\right)$, two $Q$-functions, denoted by $Q_{\beta_1}\!\left(\!\bm s_n,\bm a_n\!\right)$ and $Q_{\beta_2}\!\left(\!\bm s_n,\bm a_n\!\right)$, two target networks, denoted by $Q_{\overline\beta_1}\!\left(\!\bm s_n,\bm a_n\!\right)$ and $Q_{\overline\beta_2}\!\left(\!\bm s_n,\bm a_n\!\right)$, and an experience replay memory, where $\bm s_n$ and $\bm a_n$ are the environment state and the action in slot $n$, respectively.

\subsubsection{Policy Function}
Policy function $\pi_\phi$ performs as an “actor”. In slot $n$, the policy collects the state information from the network. The state includes a 2-dimensional UAV location, a $2M$-dimensional terminal location, an $M$-dimensional terminal battery energy, and a 1-dimensional current slot, which is formally defined by
\begin{equation}\label{eq9}
\bm s_n=\left(\mathbf q_u[n];\mathbf q_1[n],\cdots,\mathbf q_M[n];e_1[n],\cdots,e_M[n];n\right),
\end{equation}
where $e_m[n]$ is the battery energy of terminal $m$ in slot $n$. According to state $\bm s_n$, the policy function takes an action defined by a $\left(3M\!+\!2\right)$-dimensional vector
\begin{equation}\label{eq10}
\begin{split}
\bm a_n=&\big(v_u[n];\theta_u[n];p_1[n],\cdots,p_M[n];\\
&f_1[n],\cdots,f_M[n];t_1[n],\cdots,t_M[n]\big),
\end{split}
\end{equation}
to adjust the horizontal flight speed and direction of the UAV, the transmit power, CPU frequency, and offloading time proportion of each terminal.

The policy function is implemented by a deep neural network (DNN), of which the parameter is denoted by $\phi$. The DNN has two output layers. During the training process, the DNN generates the mean and covariance of a Gaussian random variable at the two output layers. Sampling the Gaussian random variable and then restricting it via a tanh function, the policy function produces an action.

The actions generated by the policy function might not meet all the constraints of problem $\mathbf P_1$. To satisfy constraints $C2$ and $C3$, we should adjust the generated actions as follows. $C2$ restricts the energy consumed by each terminal in each slot. If the generated action for a terminal does not satisfy this constraint, we set the transmit power and the CPU frequency of this terminal to zero in this slot. As a result, the computation bit of this terminal is also zero in this slot, which can be regarded as a penalty for this infeasible action. To satisfy $C3$, the offloading-time constraint, we normalize the proportion of offloading time of the generated action. Let $\hat t_m[n]$ be the proportion of offloading time of the $m$th terminal generated by the policy in slot $n$. If $\sum_{m=1}^M\hat t_m[n]\le1$, constraint $C3$ is met, thus $t_m[n]=\hat t_m[n]$; otherwise, $\hat t_m[n]$ is normalized as follows
\begin{equation}\label{eq11}
t_m[n]=\frac{1}{\sum_{m=1}^M\hat t_m[n]}\hat t_m [n].
\end{equation}
After that, SAC-TR exports the adjusted action $\bm a_n$ and obtains a reward, which is denoted by $r_n$ and will be defined in Section \ref{algorithm-reward}.

\subsubsection{Experience Replay Memory}
After getting $r_n$ and the state of the next slot $\bm s_{n+1}$ from the MEC network, SAC-TR combines $\bm s_n$, $\bm a_n$, $r_n$, and $\bm s_{n+1}$ as a sample and stores it in the experience replay memory. Once the memory is full, the newly generated sample will replace the oldest one. At fixed intervals, SAC-TR randomly selects a batch of samples from the memory and performs gradient descent on the neural networks of policy function and $Q$-functions.

\subsubsection{$Q$-function}
Following the clipped double-$Q$ trick \cite{1}, SAC-TR uses two $Q$-functions $Q_{\beta_1}$ and $Q_{\beta_2}$ as a “critic” in the gradient descent process of DNN of the policy function, such that the positive deviation of policy promotion can be reduced. $Q_{\beta_1}$ and $Q_{\beta_2}$ are performed by two DNNs with parameters $\beta_1$ and $\beta_2$. They both generate $Q$-values of a state-action pair. SAC-TR selects the small one of two $Q$-values.

\subsubsection{Target Network}
The DNN of each $Q$-function is also updated by gradient descent, where two target networks $Q_{\overline\beta_1}$ and $Q_{\overline\beta_2}$ are used to reduce the correlation between samples so as to stabilize the training. As the backup of $Q$-functions, the initial structure and the parameters of two target networks are the same as those of two $Q$-functions. They update their parameters, using the exponentially moving averages of parameters of $Q_{\beta_1}$ and $Q_{\beta_2}$, with a smoothing constant $\tau$.

\subsection{Heterogeneous Reward Function}\label{algorithm-reward}
To meet different types of requirements, including the computation rate, fairness, and specified UAV destination, we customize the reward in SAC-TR as a heterogeneous function of a computation reward and an arrival reward. In particular, we design the computation reward based on the fairness index to maximize the objective of problem $\mathbf P_1$, and design the arrival reward based on progress estimate to meet the arrival constraint $C7$.

\subsubsection{Computation Reward with Fairness}
We aim to maximize the computation bits of terminals while guaranteeing the fairness among them. On one hand, we include the incremental computation bits, i.e., $\sum_{m=1}^MU_m[n]$ in the reward to encourage the improvement of computation bits in slot $n$. On the other hand, to make use of existing information to promote fairness in each slot, we define an indicator, called current fairness index, corresponding to the definition of fairness index $I$ in (\ref{eq7}) as follows
\begin{equation}\label{eq12}
I_n=\frac{\left(\sum_{m=1}^{M}\sum_{i=1}^nU_m[i]\right)^2}{M\cdot\sum_{m=1}^{M}\left(\sum_{i=1}^nU_m[i]\right)^2}
\end{equation}
to measure the fairness among terminals in slot $n$. Eq.\ref{eq12} can be regarded as an evolution of fairness index. Clearly, there is $I_N=I$.

Combining $I_n$ with the incremental computation bits, we design the computation reward so that it can encourage actions that increase more computation bits and the actions that achieve high fairness, thereby promoting the final fairness. In the $n$th slot, the computation reward is given by
\begin{equation}\label{eq13}
r\!_c\left(\bm s_n,\bm a_n\right)=I_n^\omega\cdot\sum_{m=1}^MU_m[n],\quad n\in\mathcal{N}.
\end{equation}
\subsubsection{Arrival Reward Based on Progress Estimate}
It is important to set a proper arrival reward to facilitate UAV arriving at the designated destination at the end of the flight. Otherwise, the UAV may take a long time to (or even cannot) reach the designated destination. An example of arrival reward is the sparse reward in \cite{18}, where a fixed reward is given when the UAV arrives at the destination, or a fixed penalty when the UAV does not. However, in our problem, the area of destination is much smaller than the whole flight area, and thus the samples that the UAV arrives at the destination would be rare in the training process. As a result, if our algorithm employs the sparse reward, it will be difficult to converge.

Inspired by the progress estimate reported in \cite{24}, we design a distance-based arrival reward. The idea of the progress estimate is that, if the goal is not reached, an artificial progress estimator is given to accelerate convergence. Based on this idea, we define an arrival reward $r\!_d\left(\bm s_{N+1}\right)$ at the end of the flight according to the distance between the UAV and destination as follows
\begin{equation}\label{eq14}
r\!_d\left(\bm s_{N+1}\right)=A_1-A_2\Vert \mathbf q_u[N+1]-\mathbf q_D\Vert,
\end{equation}
where $A_1,A_2>0$ are constants. Clearly, $r\!_d\left(\bm s_{N+1}\right)$ decreases linearly with the distance between the destination and the final location of the UAV. In this way, the samples that the UAV fails to arrive at the destination can also be utilized in the training process to guide the algorithm.

Combining computation reward $r\!_c\left(\bm s_n,\bm a_n\right)$ and arrival reward $r\!_d\left(\bm s_{N+1}\right)$, we provide a heterogeneous reward function to meet multiple demands in our problem as follows.
\begin{equation}\label{eq15}
r\!\left(\bm s_n,\bm a_n\right)=\left\{
\begin{aligned}
&A_3r\!_c\left(\bm s_n,\bm a_n\right),\quad n=1,2,\cdots,N-1,\\
&A_3r\!_c\left(\bm s_n,\bm a_n\right)\!+\!r\!_d\left(\bm s_{N+1}\right), n=N,
\end{aligned}
\right.
\end{equation}
where $A_3>0$ is used to adjust the value of $r\!\left(\bm s_n,\bm a_n\right)$ to affect the convergence performance of SAC method \cite{30} and will be discussed in Section \ref{evaluation}.
\begin{algorithm}[t]
\caption{The SAC-TR algorithm} 
% \LinesNumbered
\KwIn{Randomly initialized parameters of policy function and $Q$-functions: $\phi$, $\beta_1$, and $\beta_2$;\quad Parameters of target networks: $\overline\beta_1\gets\beta_1, \overline\beta_2\gets\beta_2$;\quad\quad\quad\quad\quad\quad\quad\quad\quad\quad\quad An empty experience replay memory}
\KwOut{Optimized $\phi$, $\beta_1$ and $\beta_2$}
\While{not convergence}{
        Observe initial state $\bm s_1$\;
        \For{slot $n=1,2,\cdots,N$}{
                Sample action from policy $\bm a_n\sim\pi_\phi\left(\bm a_n|\bm s_n\right)$\;
                Execute action, collect reward $r\!\left(\!\bm s_n,\bm a_n\!\right)\!=\!r\!_c\!\left(\!\bm s_n,\bm a_n\!\right)$, observe next state $\bm s_{n\!+\!1}$\;
                \If{$n=N$}{Achieve extra reward $r\!\left(\bm s_n,\bm a_n\right)=r\!\left(\bm s_n,\bm a_n\right)+r_d\!\left(\bm s_{N+1}\right)$\;}
                Store $\left(\bm s_n,\bm a_n,r\left(\bm s_n,\bm a_n\right),\bm s_{n+1}\right)$ into memory\;
        }
        \If{it's time for an update}{
                Randomly select a sample batch from memory\;
                Update $Q$-function parameters with (\ref{eq17}):\\
                $\beta_i\gets\beta_i-\lambda\nabla_{\beta_i}J_Q\left(\beta_i\right)$, for $i\in\left\{1,2\right\}$\;
                Update policy-function parameter with (\ref{eq18}):\\
                $\phi\gets\phi-\lambda\nabla_\phi J_\pi\left(\phi\right)$\;
                Update target network:\\
                $\overline\beta_i\gets\tau\beta_i+\left(1-\tau\right)\overline\beta_i$ for $i\in\left\{1,2\right\}$\;
                Where $\lambda$ is the learning rate.
        }
}
\end{algorithm}

\subsection{Maximum Entropy Reinforcement Learning}\label{algorithm-entropy}
SAC uses the concept, called entropy of policy, to indicate the randomness of policy and is given by $\mathbb{E}_{\bm a_n\sim\pi}[-\textrm{log}\left(\pi\left(\bm a_n|\bm s_n\right)\right)]$ \cite{1}. The objective of SAC is to maximize the expectation of accumulated rewards and the expected entropy of the policy, such that the policy can be trained with various highly random samples. In this way, SAC can avoid falling into a local optimum. This objective is called the maximum entropy objective in SAC. To solve problem $\mathbf P_1$, SAC-TR defines the maximum entropy objective
\begin{equation}\label{eq16}
J(\pi)=\sum_{n=1}^N\mathbb{E}_{(\bm s_n,\bm a_n)\sim\rho_\pi}\left[r\left(\bm s_n,\bm a_n\right)-\alpha\textrm{log}\left(\pi(\cdot|\bm s_n)\right)\right].
\end{equation}
based on the reward function in (\ref{eq15}), where $\alpha$ is the temperature parameter that adjusts the importance of entropy against the reward and controls the stochasticity of policy.

At a fixed interval, SAC-TR performs gradient descent on the neural networks of $Q$-functions and the policy function. The parameters of $Q$-function $\beta_i$, $i=1,2$, are updated by minimizing the soft Bellman residual \cite{1}
\begin{equation}\label{eq17}
\begin{split}
J_Q\left(\beta_i\right)=\mathbb{E}_{(\bm s_n,\bm a_n)\sim \mathcal{D}}\Bigg[\frac{1}{2}\Bigg(Q_{\beta_i}(\bm s_n,\bm a_n)-\bigg(r(\bm s_n,\bm a_n)\\
+\gamma\Big(\!\min_{j=1,2}\!Q_{\overline\beta_j}\!(\bm s_{n+1},\bm a_{n+1})\!-\!\alpha\textrm{log}\pi_\phi(\bm a_{n+1}|\bm s_{n+1})\!\Big)\!\!\bigg)\!\!\Bigg)\!^2\!\Bigg]\!,
\end{split}
\end{equation}
where $\mathcal{D}$ is the distribution of sampled states and actions. The parameter of policy function, $\phi$, is updated by
\begin{equation}\label{eq18}
\begin{split}
J_\pi(\phi)=\mathbb{E}_{\bm s_n\sim\mathcal{D},\epsilon_n\in\mathcal{N}}\Big[\alpha\textrm{log}\pi_\phi\left(f_\phi\left(\epsilon_n;\bm s_n\right)|\bm s_n\right)\\
-\min_{j=1,2}Q_{\beta_j}\left(\bm s_n,f_\phi\left(\epsilon_n;\bm s_n\right)\right)\Big].
\end{split}
\end{equation}
In (\ref{eq18}), the reparameterization trick is employed as the solution for policy gradient \cite{1}, in which the policy is rewritten as $\bm a_n=f_\phi(\epsilon_n;\bm s_n)$, where $\epsilon_n$ is an independent noise vector, as shown in Fig.\ref{fig3}. 

Before use, SAC-TR is trained until it converges, of which the training process is summarized by Algorithm 1. The well-trained algorithm is then carried by the UAV as an agent. At the beginning of each slot, the UAV collects the state information and makes a decision. During the flight time, SAC-TR can continue to be trained at a fixed interval if needed.

\section{Performance Evaluation}\label{evaluation}
In this section, we evaluate the performance of SAC-TR by simulations. In particular, we study the convergence, usability, and adaptability of SAC-TR, the effect of the exponent of the fairness index, and the optimal policy given by SAC-TR. We also compare SAC-TR with other benchmarks. 

\subsection{Simulation Settings}\label{evaluation-setting}
\subsubsection{System Settings}
In the simulation, we set the total flight time $T\!=\!4$ seconds, which is discretized into $N\!=\!40$ slots, and the number of mobile terminals $M\!=\!4$. The maximum flight speed of UAV $v_u^{max}\!=\!30$ m/s, the data offloading bandwidth $B\!=\!40$ MHz, the carrier frequency $f_c\!=\!2.4$ GHz, and the receiver noise power $\sigma^2\!=\!{10}^{-9}$ Watts \cite{5}. The WPT energy conversion efficiency at each terminal $\eta_0\!=\!0.8$ \cite{34}. The effective capacitance coefficient of the terminal $\zeta_c\!=\!{10}^{-28}$, which depends on the chip architecture, and the CPU cycle of raw data $C\!=\!100$ cycles/bit \cite{21,37}. The upload data needed for each bit of raw data $\delta\!=\!1$. In remote area, the parameters $\left(h,l,\eta_{LoS},\eta_{NLoS}\right)$ in (\ref{eq2}) are $\left(4.88,0.43,0.1,21\right)$ \cite{11}. A field with a horizontal area of $18\times18$ m$^2$ is considered, and the flight altitude of UAV $H\!=\!5$ m. The horizontal location of the starting point of the UAV is $\left(0,0\right)$ m, and the destination range is a sector with the center of $\left(18,18\right)$ m and the radius of $1$ m, as shown in Fig. \ref{fig10}.
\begin{figure}[!t]
\centering
\includegraphics[width=3in]{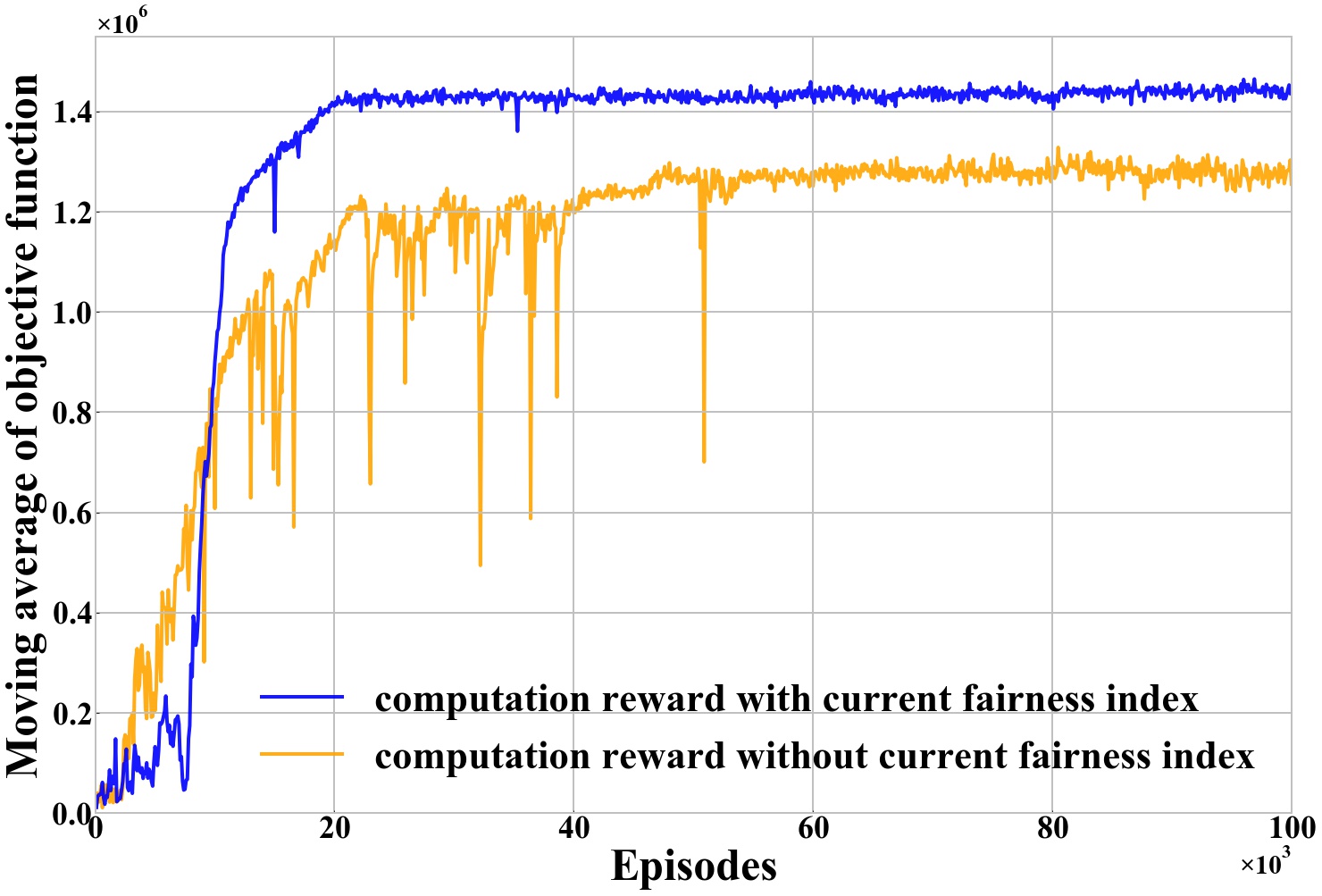}
\caption{Moving average of objective function under different designs of computation reward.}
\label{fig4}
\end{figure}

\subsubsection{Mobility Model of Terminals}
Since the mobility model of the terminal may contain fixed components, randomness, and memory, we employ the Gauss-Markov random model (GMRM) to characterize it \cite{15}. Assume the speed and the direction of the $m$th terminal in the $n$th slot are respectively $v_m[n]$ and $\theta_m[n]$, they can be calculated in GMRM by
\begin{subequations}\label{eq19}
\begin{align}
v_m[n]\!=\!k_{1,m}v_m[n\!-\!1]\!+\!\left(\!1\!-\!k_{1,m}\!\right)\overline v_m\!+\!\!\sqrt{1\!-\!k_{1,m}^2}\Phi_m,\\
\theta_m[n]\!=\!k_{2,m}\theta_m[n\!-\!1]\!+\!\left(\!1\!-\!k_{2,m}\!\right)\overline \theta_m\!+\!\!\sqrt{1\!-\!k_{2,m}^2}\Psi_m,
\end{align}
\end{subequations}
Herein, $0\le k_{1,m},k_{2,m}\le1$ represent the memory in the mobility model of the $m$th terminal. $\overline v_m$ and $\overline\theta_m$ are the average speed and average direction of the $m$th terminal. $\Phi_m$ and $\Psi_m$ are Gaussian distributed random variables, which inflect the randomness in the mobility model of the $m$th terminal. In the simulation, we set $\overline\theta_1\!=\!\overline\theta_3\!=\!0$, $\overline\theta_2\!=\!\overline\theta_4\!=\!\pi$, and $\overline v_m\!=\!2$ m/s, $k_{1,m}\!=\!k_{2,m}\!=\!0.9$, the mean and covariance of $\Phi_m$ as 0 and 2, and that of $\Psi_m$ as 0 and 1, for $m\in \mathcal{M}$. Note that SAC-TR can also be applied to other mobility models of terminals, including changeable or unknown models.
\subsubsection{Simulation Platform}
We execute SAC-TR in Python 3.7 with PyTorch 1.7. The neural networks in policy function and Q-functions are both fully connected networks, each of which has three hidden layers and each hidden layer has 400 neurons. We adopt the Adam optimizer and utilize the RELU as the activation function. We set the discount $\gamma\!=\!0.8$ and the algorithm is updated every 100 slots. The parameters $A_1$ and $A_2$ in arrival reward are respectively set as 500 and 80. It is pointed out in \cite{30} that the SAC method performs well when the average reward in each slot is around dozens. Thus, we regulate the average reward in this range by setting different $A_3$ in different scenarios. For example, when the exponent of the fairness index $\omega\!=\!4$ and the transmit power of the UAV $P_e\!=\!0.1$ Watts, we select $A_3\!=\!4.9\times{10}^{-4}$ to make the return to be around 1200 (30 per slot) after convergence. Herein, the return is the accumulated reward in an episode, and the episode is an independent realization of an entire flight time. 

\subsection{Convergence of SAC-TR}\label{evaluation-convergence}
\begin{figure}[!t]
\centering
\subfigure[]{
 \label{fig5a}
 \includegraphics[width=2.8in]{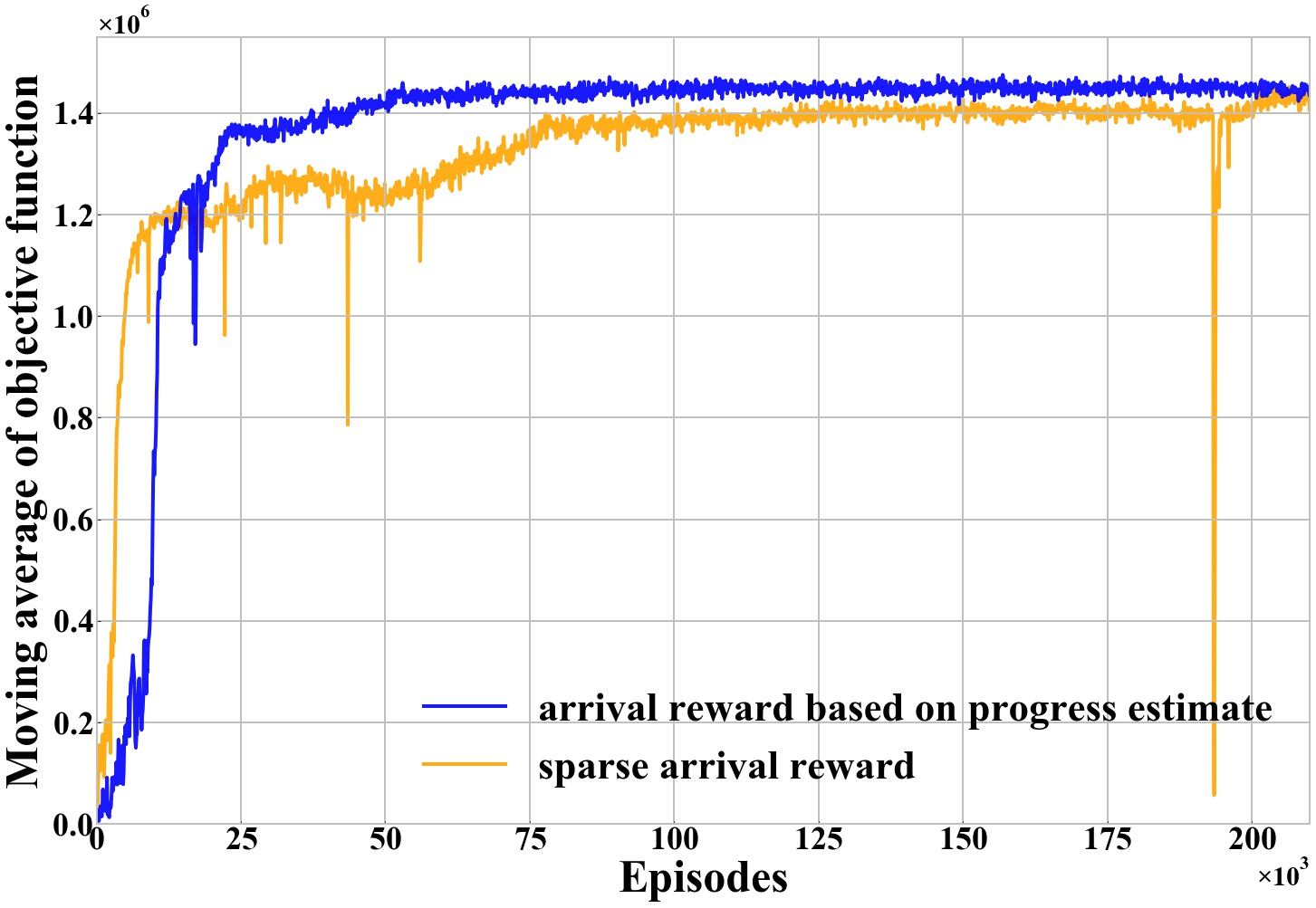}}
\subfigure[]{{}
 \label{fig5b}
 \includegraphics[width=2.8in]{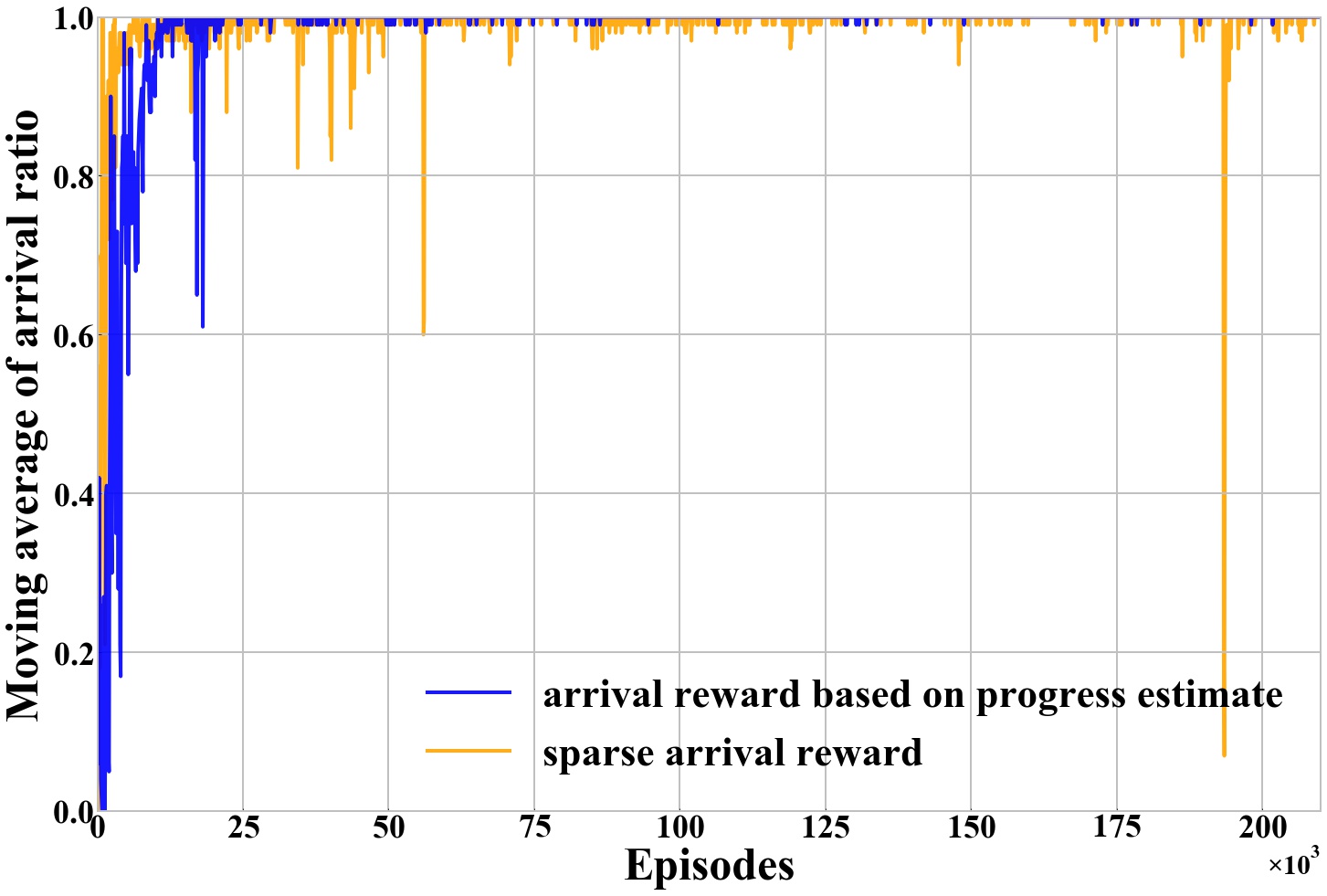}}
\caption{Moving average of (a) objective function and (b) arrival ratio under different designs of arrival rewards.}
\label{fig5}
\end{figure}
In this part, we study the effect of the reward function and hyperparameters on the convergence of SAC-TR, where we set $P_e=0.1$ Watts and $\omega=4$.

The current fairness index $I_n$ is integrated into the computation reward to improve fairness. To investigate its effect, we compared SAC-TR to that without $I_n$ in the computation reward, as plotted in Fig.\ref{fig4}. The curves denote the moving average of the objective function over a window of 100 episodes. As we can see, the computation reward integrated with $I_n$ reaches higher objective function compared to that without $I_n$, since $I_n$ can guarantee fairness in almost all the situations.

We also design an arrival reward based on the progress estimate to promote convergence. To examine its effect, we compare SAC-TR to that with sparse arrival reward. With sparse arrival reward, if the UAV reaches the destination, a fixed reward $A_1$ will be given, otherwise, no arrival reward will be given. Fig.\ref{fig5} shows the moving average of the objective function and arrival ratio under different arrival rewards. Herein, the arrival ratio is the ratio that the UAV arrives at the destination successfully over the last 100 episodes. Fig.\ref{fig5a} shows that the objective function converges to a stable value around the $70,000$th episode under the reward based on the progress estimate, while it takes $210,000$ episodes to reach the same value under the sparse arrival reward. Fig.\ref{fig5b} shows that we achieve the goal of reaching the destination more quickly and stably under the progress estimate-based reward.
\begin{figure}[!t]
\centering
\includegraphics[width=3in]{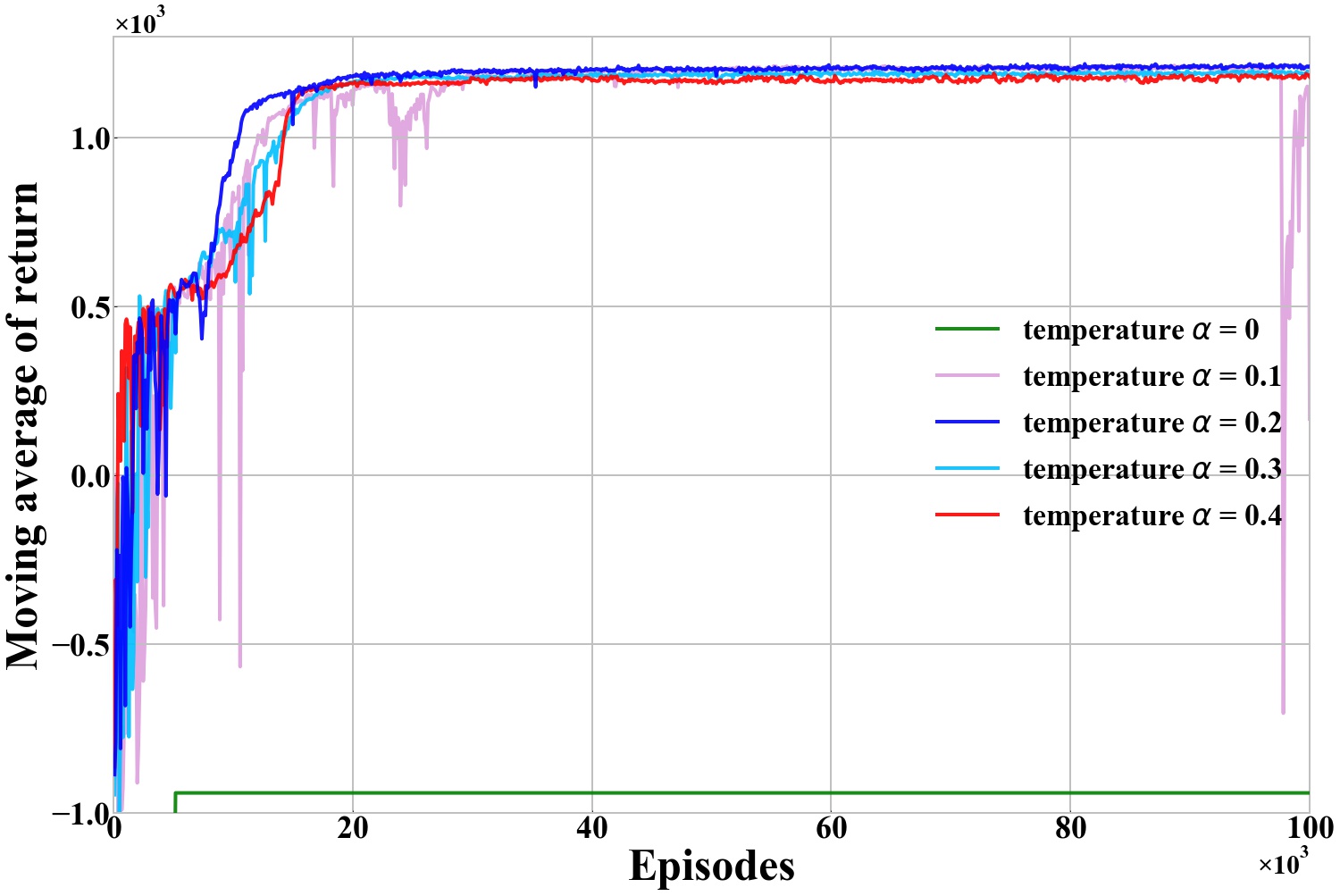}
\caption{Moving average of return under different temperature parameters $\alpha$.}
\label{fig7}
\end{figure}

\begin{figure*}[!t]
\centering
\subfigure[]{\includegraphics[width=2.8in]{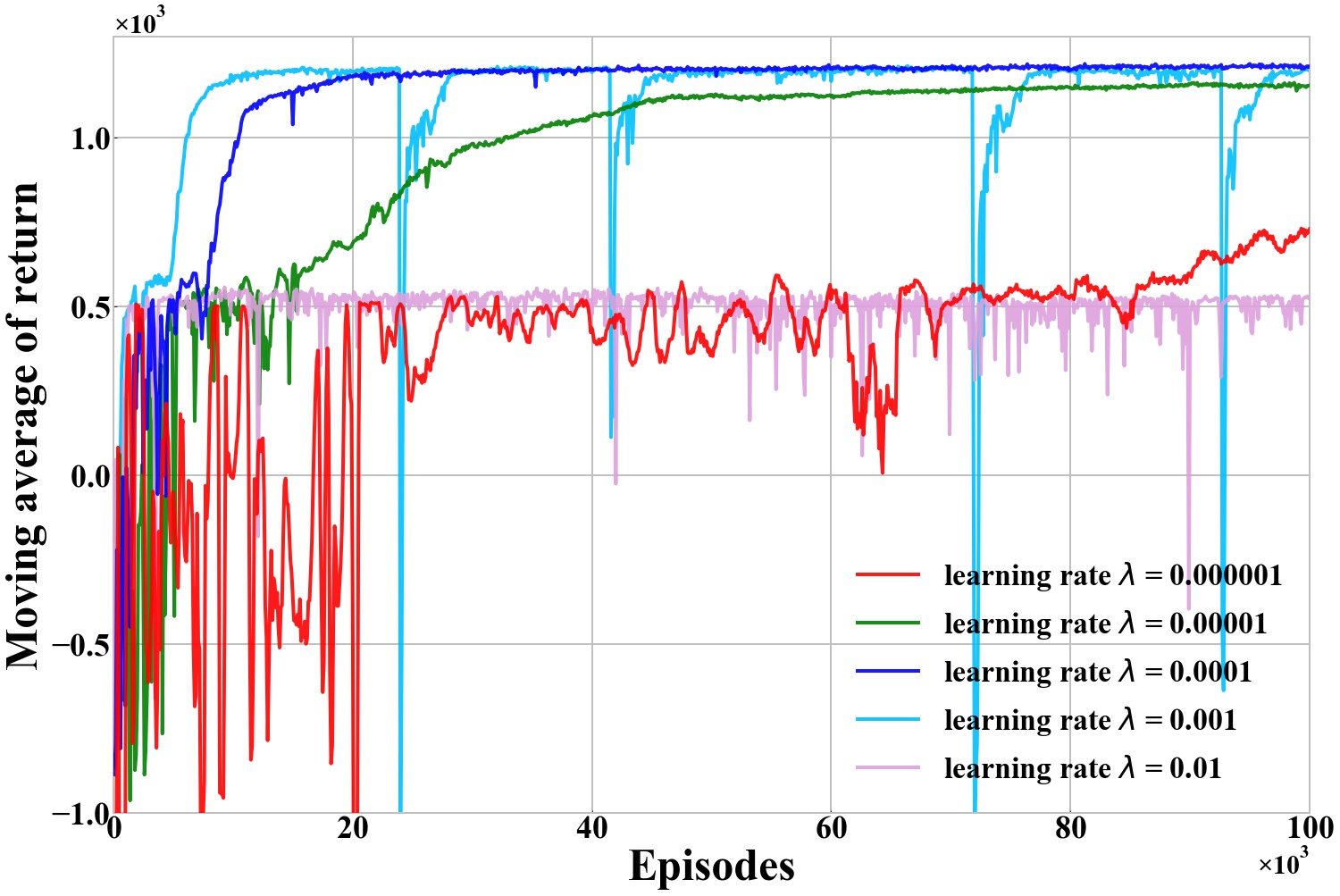}
\label{fig8a}}
\hfil %as a separator to get equal spacing
\subfigure[]{\includegraphics[width=2.8in]{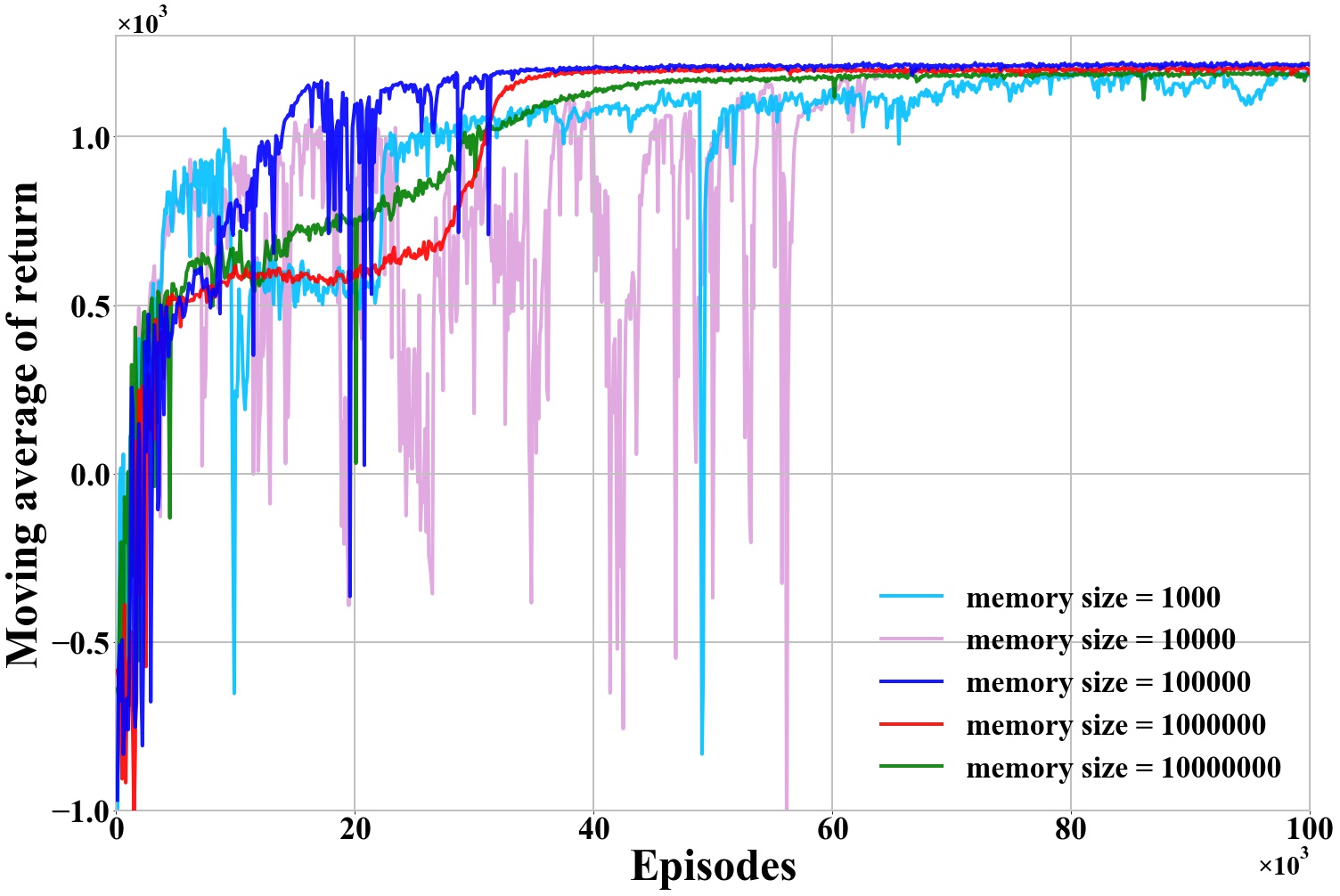}
\label{fig8b}}
\subfigure[]{\includegraphics[width=2.8in]{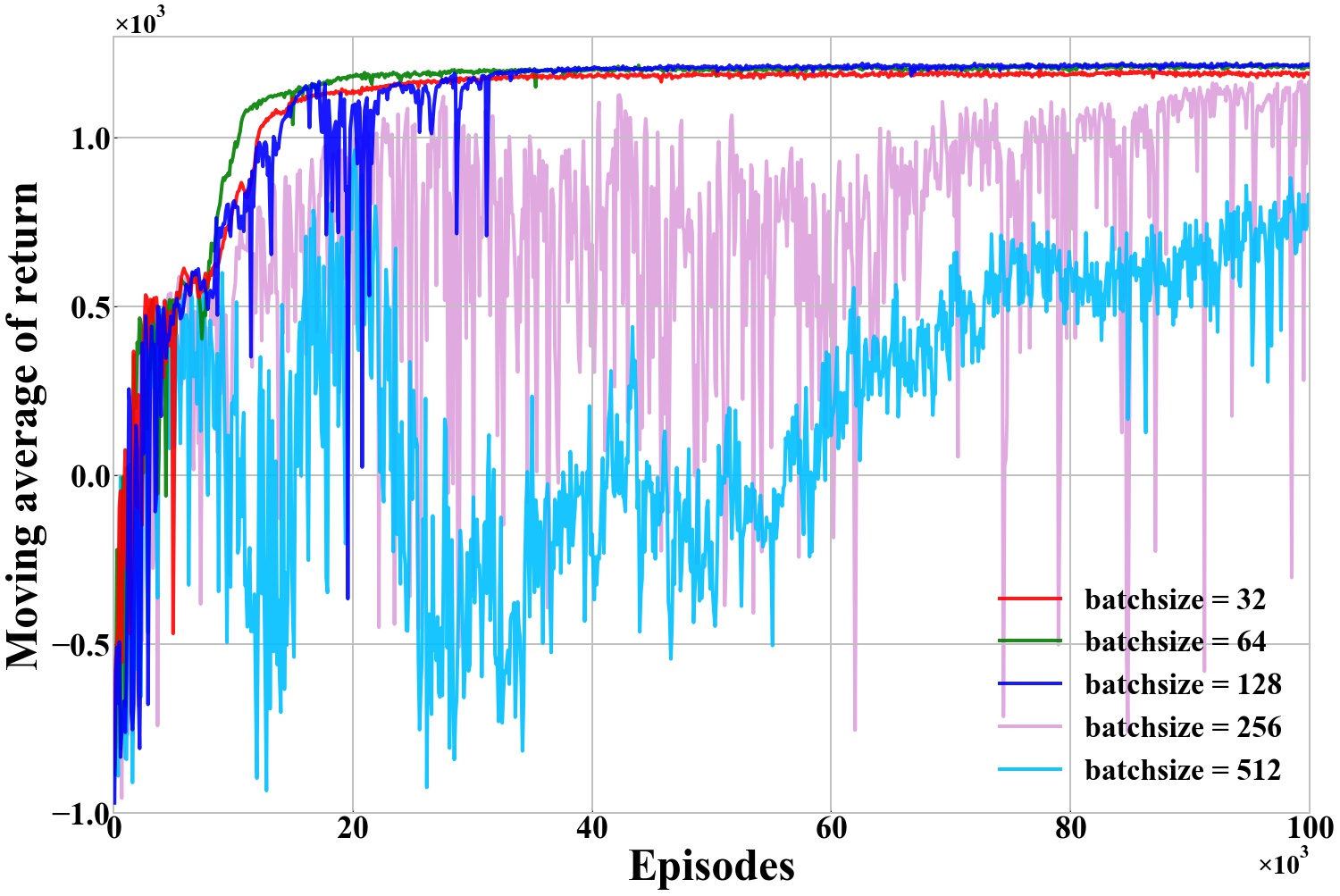}
\label{fig8c}}
\hfil %as a separator to get equal spacing
\subfigure[]{\includegraphics[width=2.8in]{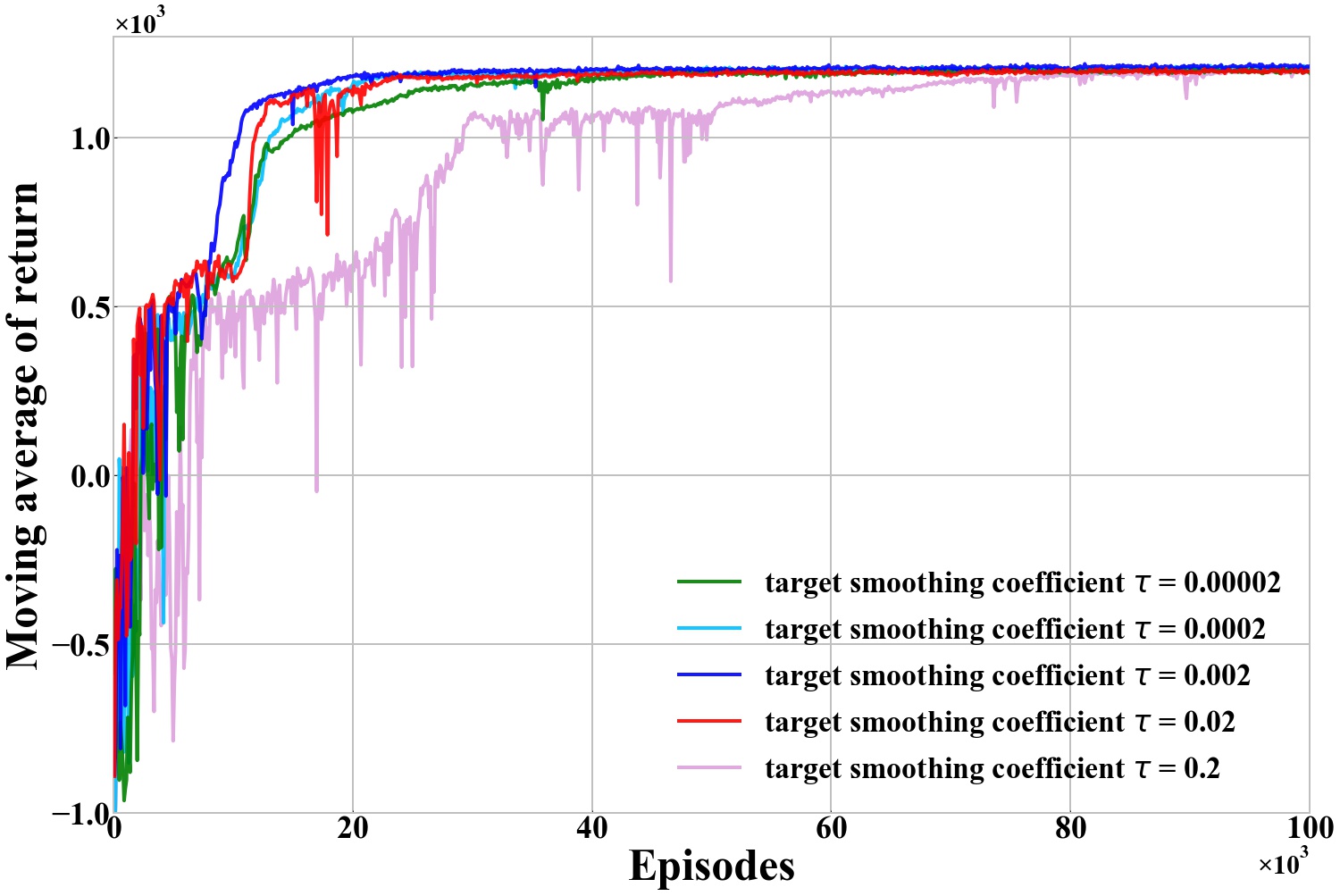}
\label{fig8d}}
\caption{Moving average of return under different algorithm hyperparameters: (a) learning rate; (b) memory size; (c) training batch size; (d) target smoothing coefficient.}
\label{fig8}
\end{figure*}
In Fig.\ref{fig7}, we investigate the effect of the temperature parameter $\alpha$ on convergence, which is utilized to control the randomness of the policy. It is shown that the return can be barely improved when $\alpha=0$. This is because the policy entropy is not included in the object of SAC-TR at this time, which causes a low exploration of the algorithm. This informs us SAC-TR cannot be fully upgraded when it only has the structure of the AC method but without policy entropy in the DRL objective. In addition, a quite large $\alpha$ (=0.4) can also lead to a local optimum. This is because a too large $\alpha$ makes the algorithm excessively pursue the improvement of randomness instead of the accumulated reward. Thus, we set $\alpha=0.2$ in the following simulations.

In Fig.\ref{fig8}, we show the experience of setting hyperparameters. Fig.\ref{fig8a} plots the moving average of the return under different learning rates $\lambda$ in the Adam optimizer. We observe that when $\lambda$ is quite large (= 0.01), the algorithm converges to a local optimum, while a small $\lambda$ stabilizes and decelerates convergence. Trading between performance and convergence speed, we set $\lambda=0.0001$ in the following simulations. At a fixed interval, SAC-TR randomly selects a batch of samples from the experience replay memory to train neural networks. In Fig.\ref{fig8b} and \ref{fig8c}, we respectively study the effect of the size of the experience replay memory and the training batch size on convergence. Fig.\ref{fig8b} indicates that either too small or too large memory size decelerates convergence. Therefore, we select memory size of 100,000. Fig.\ref{fig8c} shows that the algorithm may converge to a local optimum when batch size is too small (= 32), while the convergence speed slows down when the batch size is quite large ($\ge$ 256). To maximize the convergence speed, we set batch size = 64 or 128 in simulations under different cases. In SAC-TR, $Q$-functions update the values of target networks by using the exponentially moving average with a parameter called target smoothing coefficient, $\tau$. In Fig.\ref{fig8d}, we study the effect of $\tau$ on convergence. It shows that a too large $\tau$ decelerates convergence, since an overly fast update of the target network destabilizes convergence, while a too-small $\tau$ also reduces the convergence speed. For fast convergence, we set $\tau=0.002$ in the following simulations.

\subsection{Effect of Fairness Index Exponent $\omega$}\label{evaluation-exponent}
\begin{figure}[!t]
\centering
\includegraphics[width=3in]{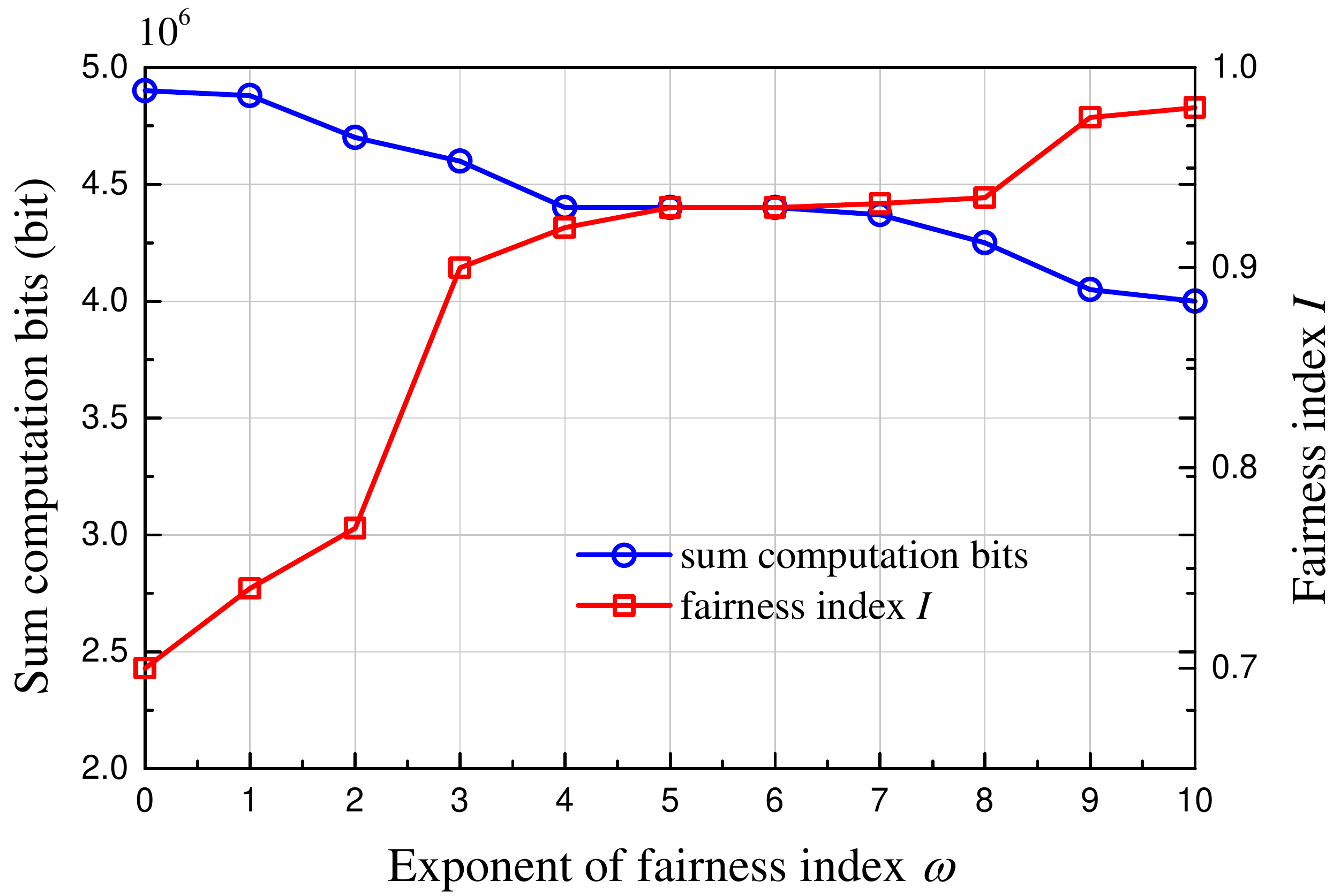}
\caption{Sum computation bits and fairness index $I$ versus the exponent of fairness index $\omega$.}
\label{fig9}
\end{figure}

In Fig.\ref{fig9}, we investigate the effect of the exponent of fairness index, $\omega$, on the sum computation bits of all terminals and the fairness index $I$ when $P_e=1$ Watts, since it is used to adjust the importance of fairness in the objective function. With the increase of $\omega$, the sum computation bits declines while $I$ monotonously increases. It indicates that the preference of SAC-TR for sum computation bits and fairness can be effectively adjusted by $\omega$. A large $\omega$ can be set when quite high fairness is demanded. In contrast, if the scenario aims to maximize the sum computation bits, we should have $\omega=0$.

\subsection{Optimal Policy}\label{evaluation-optimal}
\begin{figure}[!t]
\centering
\subfigure[]{
 \label{fig10a}
 \includegraphics[width=3in]{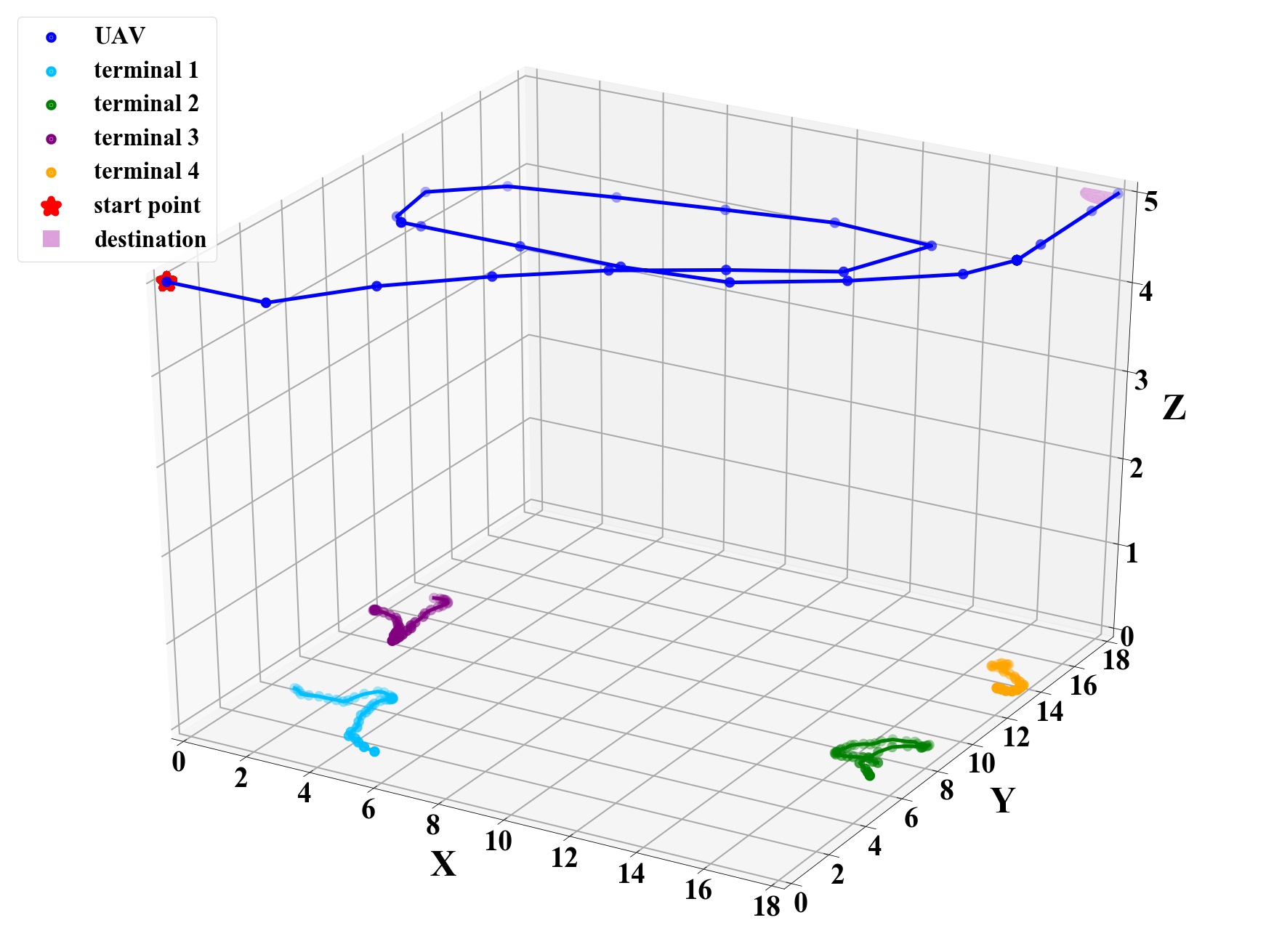}}
\subfigure[]{{}
 \label{fig10b}
 \includegraphics[width=3in]{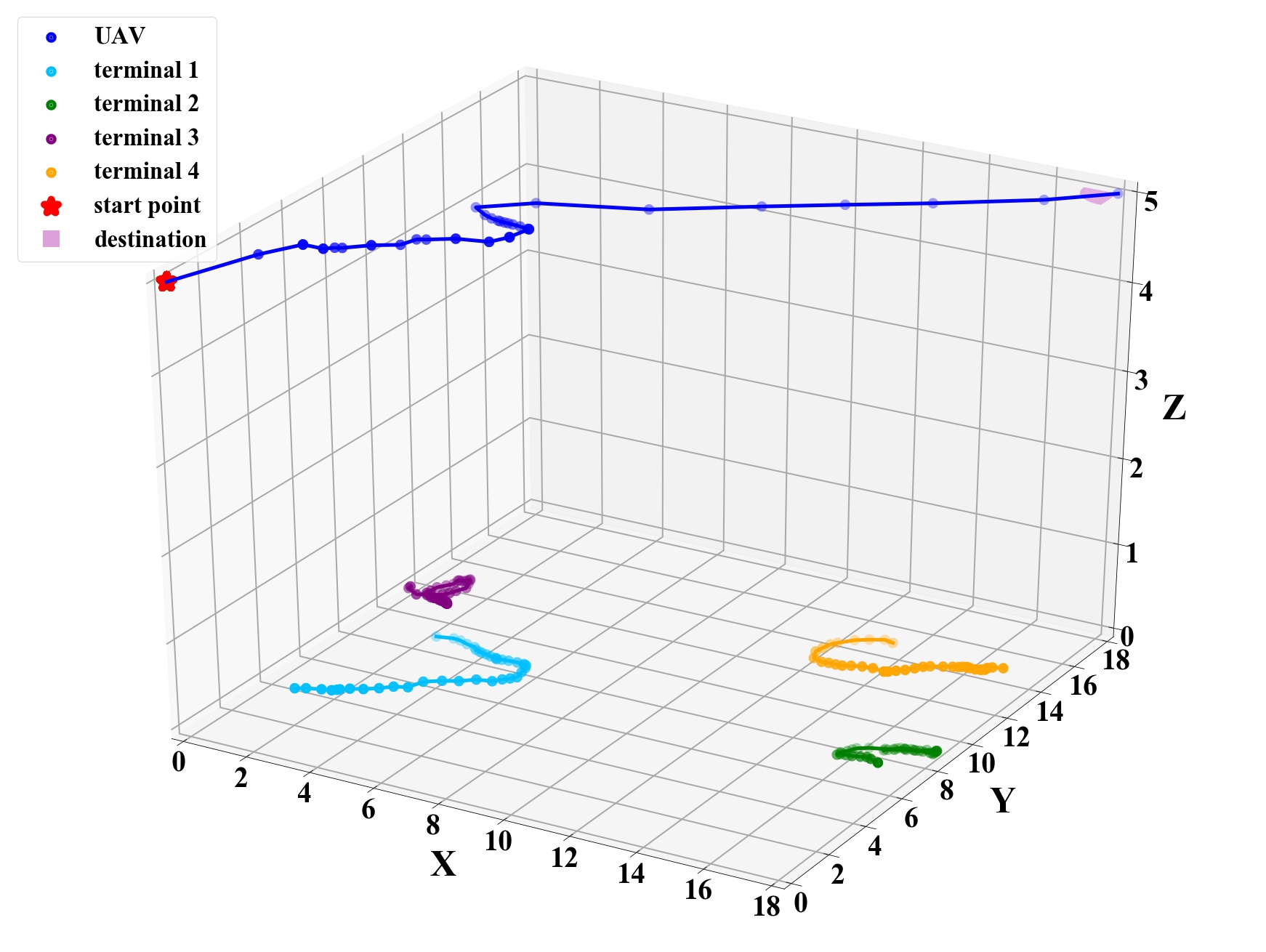}}
\caption{Examples of the optimal UAV trajectory when (a) $\omega=4$ (b) $\omega=0$.}
\label{fig10}
\end{figure}

The optimal policy given by SAC-TR includes the optimal UAV trajectory and the optimal resource allocation of terminals. Fig.\ref{fig10a} shows an example of optimal UAV trajectory, where $P_e=1$ Watts and $\omega=4$. In the entire fight time, the UAV first hovers over four mobile terminals, which can ensure the fairness among terminals, then flies towards the destination at high speed during the last few slots and finally arrives at the destination.

Fig.\ref{fig10b} plots the UAV trajectory when $\omega=0$. Unlike the case of $\omega=4$, the UAV first flies quickly to terminal 1, then adjusts its trajectory to keep itself always on the top of terminal 1, and finally reaches the destination. The difference is caused by the fact that when $\omega=0$, SAC-TR aims to maximize the sum computation bits. Therefore, the algorithm minimizes the distance between the UAV and one of the terminals to enhance communication and energy harvesting. It reveals that when $\omega=0$, SAC-TR cannot guarantee the fairness issue and an unfair phenomenon may appear such as the situation in Fig.\ref{fig10b}.

\begin{figure}[!t]
\centering
\includegraphics[width=3.5in]{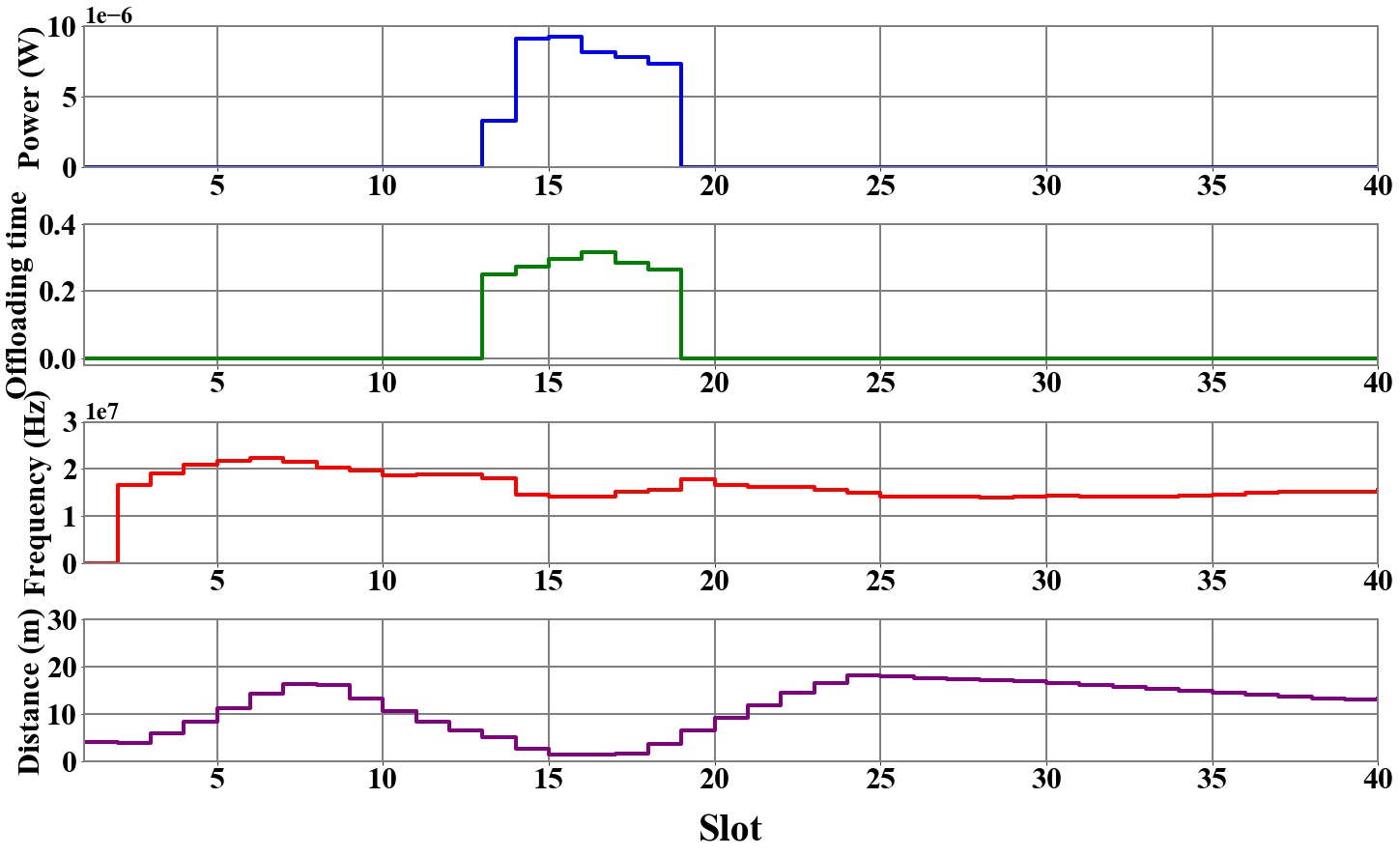}
\caption{An Example of optimal resource allocation when $\omega=4$.}
\label{fig11}
\end{figure}

Fig.\ref{fig11} plots an example of the optimal resource allocation of terminal 1 when $\omega=4$. From top to bottom, Fig.\ref{fig11} exhibits the transmit power, the proportion of offloading time, the CPU frequency, and the distance between the UAV and the terminal 1 in an entire flight time. We observe that the resource allocation of the terminal cooperates with the UAV trajectory for good performance. In particular, when the UAV flies near the terminal 1 (13 $\le$ slots $\le$ 19), the transmit power and the proportion of offloading time of this terminal increase so as to utilize the high power channel gain at this moment to offload more computation tasks. The CPU frequency is relatively uniform over the entire flight time.

\subsection{Comparison with Benchmark Algorithms}\label{evaluation-comparison}
To evaluate the performance of SAC-TR, we compare it with other representative benchmarks as follows.
\begin{enumerate}[]
\item Hover–fly–hover (HFH) trajectory algorithm. The HFH trajectory of UAV is widely used to serve the terminals with fixed locations \cite{7,29,23}. In HFH, the UAV flies to and hovers over some specified locations in turn. To serve mobile terminals, we make simple adaptations to HFH: the UAV flies to and follows each mobile terminal in turn for equal time. During following one terminal, it keeps itself on the top of this terminal in each slot. After serving one terminal, It flies to the next terminal with the maximum speed. Also, the UAV reserves the minimum time to arrive at the destination. 
\item Straight trajectory algorithm. In this algorithm, the UAV flies straight from the starting point to the destination with constant speed. 
\item Greedy local algorithm. In this algorithm, the terminal exhausts all the energy in the battery for local computation in each slot. 
\item Greedy offloading algorithm. The terminal spends all the energy in the battery for computation offloading in each slot, while the offloading time of each terminal is equal. 
\item Random algorithm. In this algorithm, the flight speed and direction of the UAV, transmit power, offloading time, and CPU frequency of each terminal in each slot are picked randomly.
\end{enumerate}
To compare with SAC-TR, the resource allocation in the HFH trajectory algorithm and the straight trajectory algorithm is optimized by our algorithm. Also, the UAV trajectory in the greedy local algorithm and the greedy offloading algorithm is optimized by our algorithm.
\begin{figure}[!t]
\centering
\subfigure[]{\includegraphics[width=1.67in]{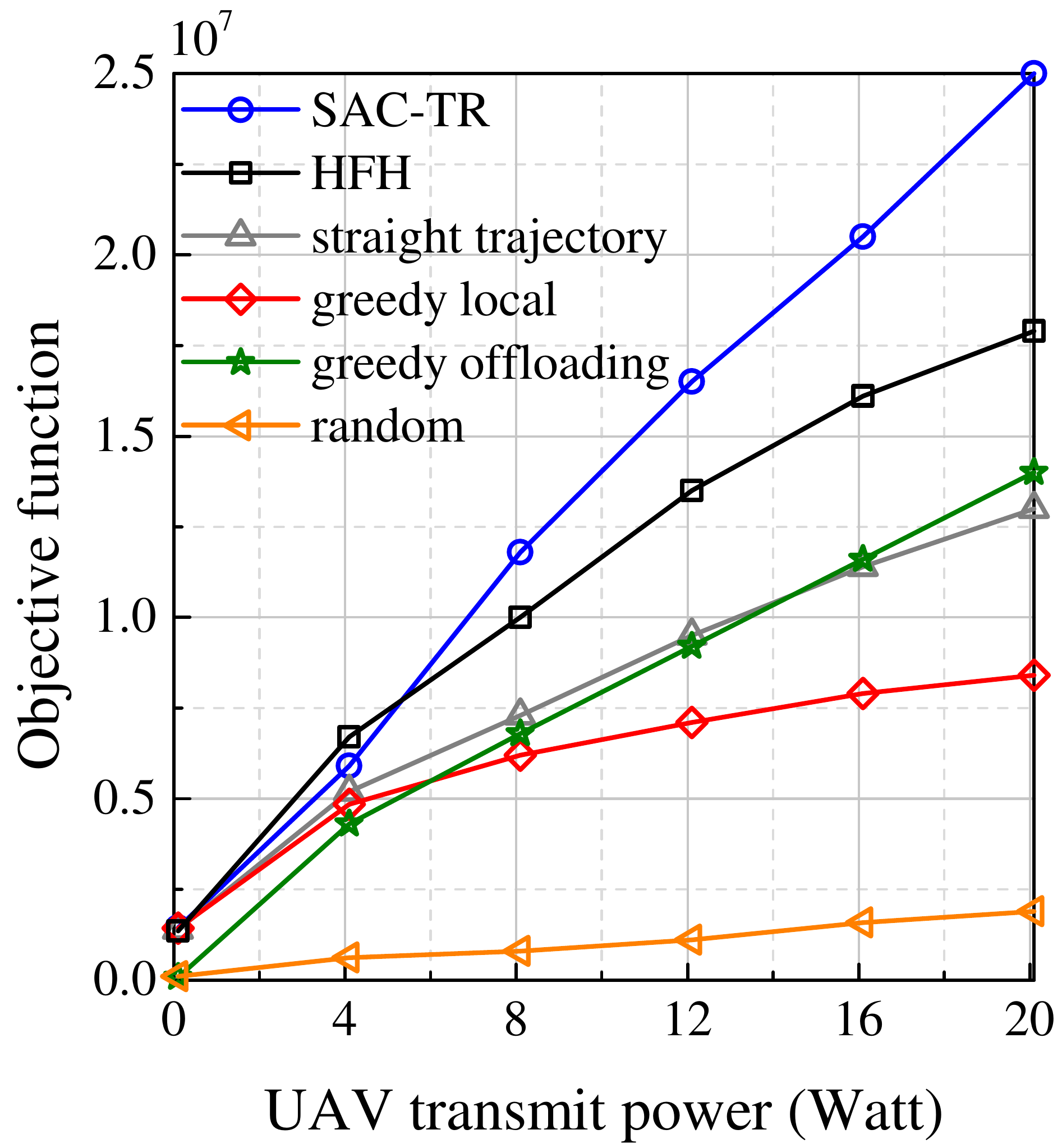}
\label{fig12a}}
% \hfil %as a separator to get equal spacing
\subfigure[]{\includegraphics[width=1.67in]{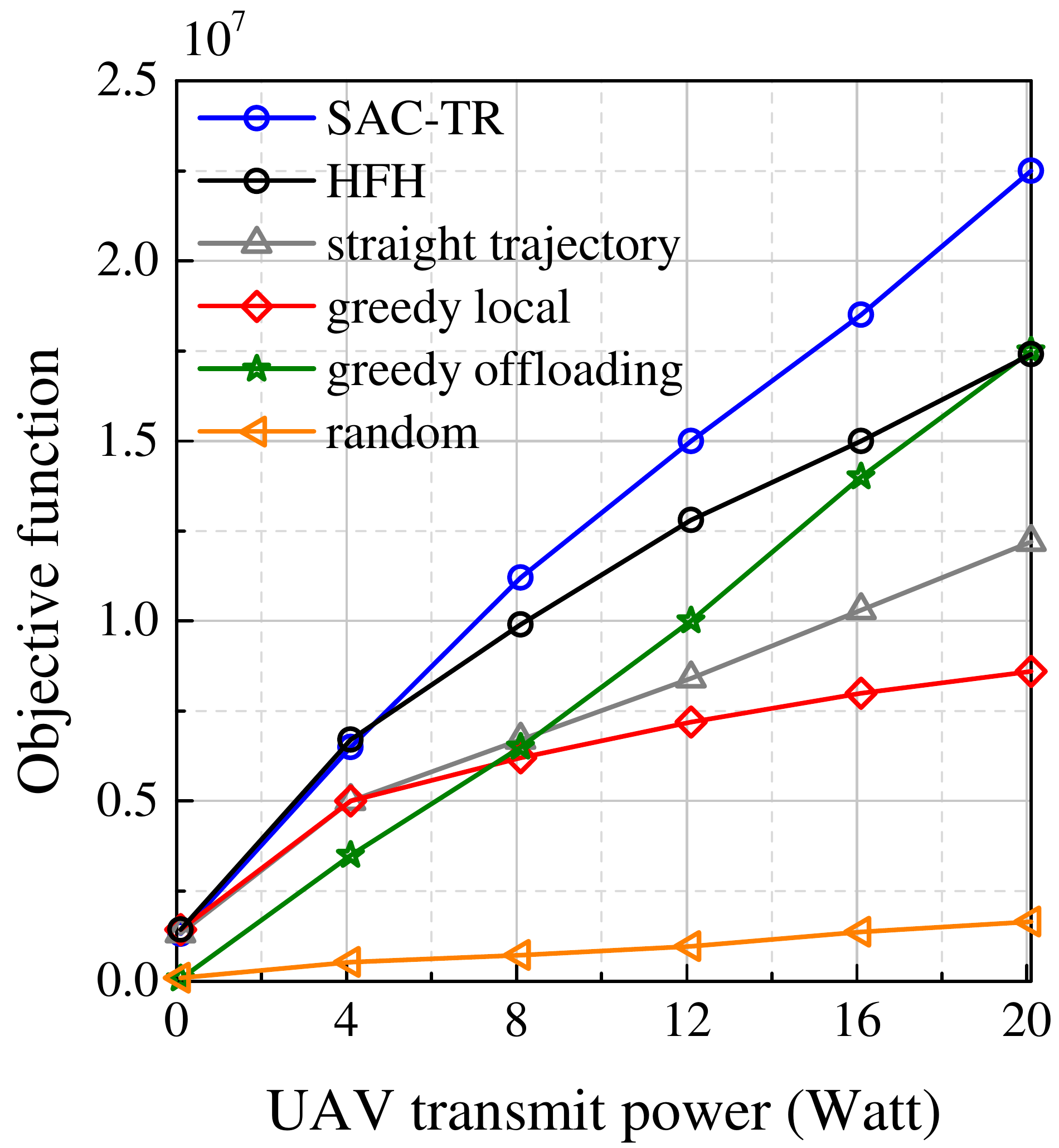}
\label{fig12b}}
\caption{The objective function of SAC-TR and other benchmarks when the average speed and the memory of each terminal (a) are the same (b) have a greater degree of difference.}
\label{fig12}
\end{figure}

In Fig.\ref{fig12a}, we investigate the stable value of the objective function after convergence, under SAC-TR and the above-mentioned algorithms. We set $P_e$ from 0.1 to 21.1 Watts and $\omega=4$. With the increase of $P_e$, mobile terminals can harvest more energy to use in local computation or computation offloading, thus the objective functions of all algorithms improve. The greedy local algorithm is better than the greedy offloading algorithm when $P_e$ is small; otherwise, the greedy offloading algorithm is better. The objective function of the HFH trajectory algorithm is always high since the HFH trajectory guarantees high fairness while keeping the distance between the UAV and terminals as small as possible to maximize the computation bits. The objective function of SAC-TR is always close to or exceeds other benchmarks. Also, SAC-TR is sometimes worse than that of the HFH trajectory algorithm, which is due to the error caused by the inconsistency of the objective function of problem $\mathbf P_1$ and the objective of SAC-TR.

Different from that in Fig.\ref{fig12a}, we consider the situation that there exists a greater degree of difference in the mobility of each terminal in Fig.\ref{fig12b}. In particular, the average speed of each terminal is different, that is, $\left(\overline v_1,\overline v_2,\overline v_3,\overline v_4\right)=\left(0.6,1.2,1.8,2.4\right)$ m/s, and so as the memory of each terminal, $\left(k_{1,1},k_{1,2},k_{1,3},k_{1,4}\right)=\left(1.0,0.8,0.6,0.4\right)$, and $k_{2,m}\!=\!k_{1,m}$, for $m\in\mathcal{M}$. As Fig.\ref{fig12b} shows, the objective function of SAC-TR still approaches or exceeds other benchmarks in this situation. We observe that, the advantage of the trajectory planning of our algorithm becomes more prominent in this situation. For example, the superiority of SAC-TR and the greedy offloading algorithm whose UAV trajectory is optimized by our algorithm is more evident compared to that in Fig.\ref{fig12a}. This is because the mobility model of each terminal is quite different in this situation, and the UAV can approach each terminal with flexible time and flexible trajectory under the trajectory planning of our algorithm, while HFH still allocates equal flight time to follow each terminal with inflexible trajectory and the straight trajectory algorithm has completely fixed UAV trajectory. 

Fig.\ref{fig12a} and \ref{fig12b} reveal the advantage of trajectory planning of SAC-TR by comparing it with HFH and straight trajectories. Also, the superiority of resource allocation of SAC-TR is highlighted when comparing it with the greedy local algorithm and the greedy offloading algorithm, of which the reason is that SAC-TR makes overall arrangements over the entire flight time. Since benchmarks (1)$\sim$(4) only perform optimization on partial parameters by our algorithm, these comparisons reflect the benefit of joint optimization of all parameters in SAC-TR.

\subsection{Usability and Adaptability}\label{evaluation-adaptability}
\begin{table} [t] 
\centering
\caption{Execution and Update Latency of SAC-TR (s)}  
\begin{tabular}{cccccc}
\toprule
Terminal number & 2 & 4 & 8 & 16 & 32 \\
\midrule
Execution latency ($\times {10}^4$) & 4.13 & 4.30 & 4.35 & 4.49 & 5.07\\
\midrule
Update latency ($\times {10}^2$) & 5.10 & 5.23 & 5.27 & 5.30 & 5.85\\
\bottomrule
\end{tabular}\label{Tb1}
\end{table}
Finally, we evaluate the usability and adaptability of SAC-TR by simulations, where we set $P_e=0.1$ Watts and $\omega=4$.

Since SAC-TR is executed at the beginning of each slot, its execution latency has a big impact on its usability. Therefore, we evaluate the execution latency on a desktop configured with an Intel Core i5-4590 3.3GHz CPU and 8 GB of memory and exhibit the results in Table \ref{Tb1}. The single execution latency is averaged over $1.6\times{10}^6$ executions. Under different numbers of mobile terminals, the execution latency is always around $5\times{10}^{-4}$s, which is much lower than the length of a slot, such as 0.1s, so its impact on performance can be ignored.

In some situations, we proceed to train SAC-TR during use with a fixed interval so as to adapt to unpredictable changes of the environment. Therefore, we also evaluate the latency of updating SAC-TR for once in table \ref{Tb1}, which is averaged over $1.6\times{10}^6$ updates with a batch size of 64. Under different numbers of terminals, the update latency is around $5\times{10}^{-2}$s. It follows that SAC-TR should be updated with an interval that is larger than the update latency. During use, SAC-TR can be trained on the CPU of the UAV or the cloud server by transmitting data to it via satellite communication \cite{35}. 
\begin{figure}[!t]
\centering
\includegraphics[width=2.8in]{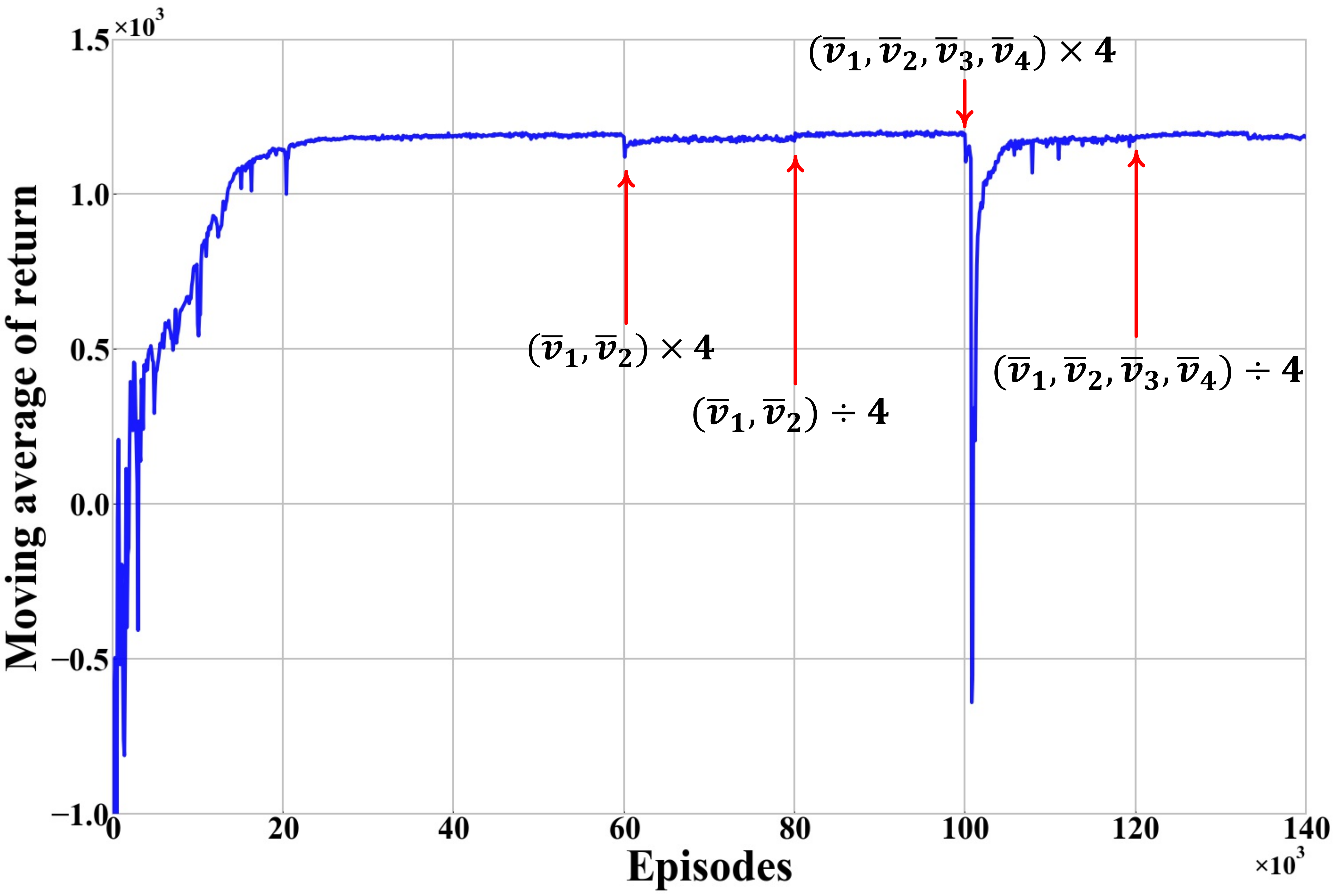}
\caption{The moving average of return when the average speeds $\overline v_m$ of terminals have large-scale changes.}
\label{fig13}
\end{figure}

On the other hand, we examine the adaptability of SAC-TR to the unexpected changes of the mobility model of terminals. In Fig.\ref{fig13}, we investigate the large-scale changes of the average speed $\overline v_m$ of terminals. In the $60,000$th slot, $\overline v_m$ of two terminals are both quadrupled, and then reduced $1/4$ to origin value in the $80,000$th slot. As we can see, SAC-TR can adapt to these abrupt changes instantly. In the $10,000$th slot, $\overline v_m$ of all terminals are quadrupled, and the performance of SAC-TR drops drastically in an instant but converges to a stable value again quickly. In the $120,000$th slot, the average speed of all terminals drops 4 times and SAC-TR handles it smoothly.
\begin{figure}[!t]
\centering
\includegraphics[width=2.8in]{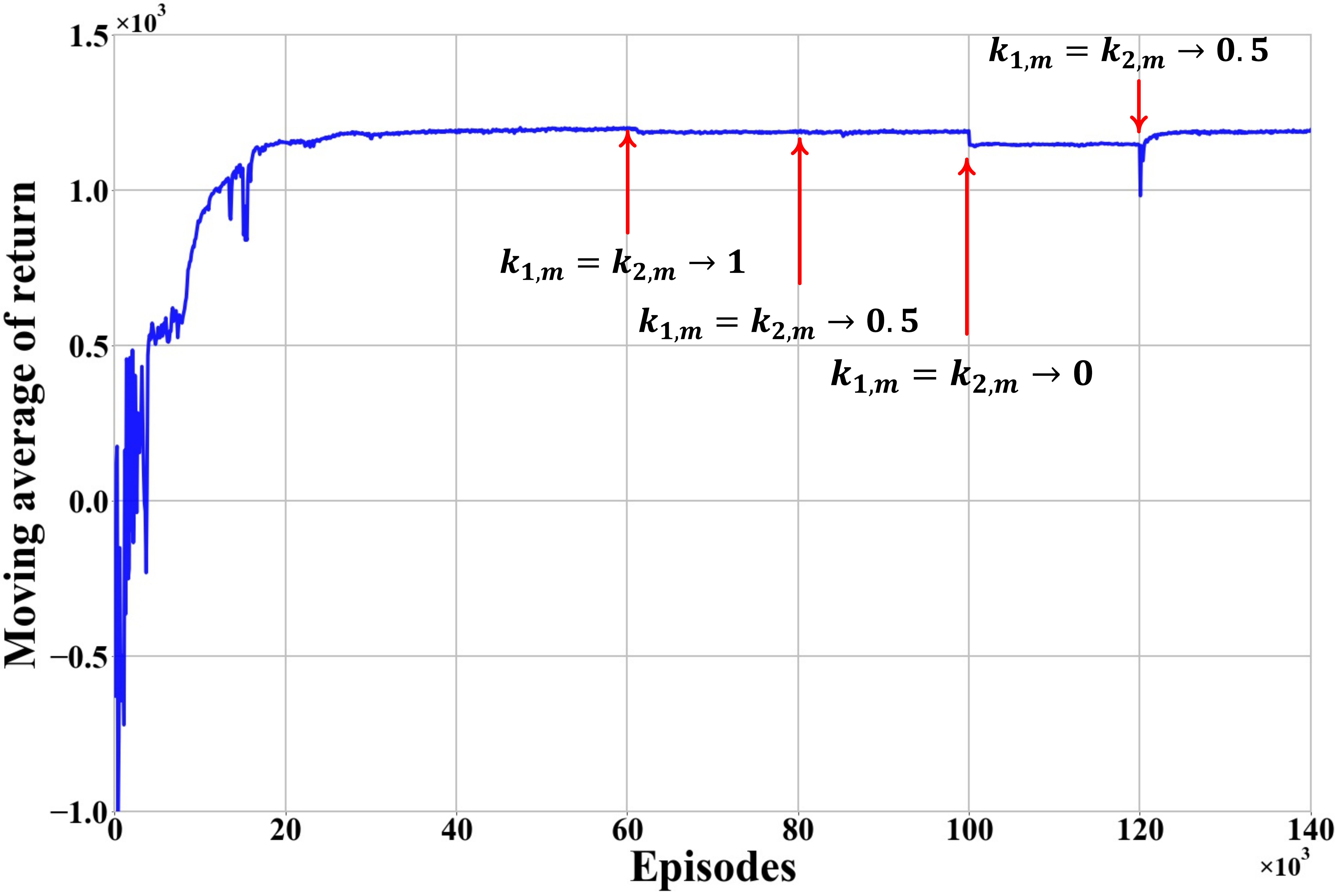}
\caption{The moving average of return when $k_{1,m}$ and $k_{2,m}$ which indicate the memory of terminals have large-scale changes.}
\label{fig14}
\end{figure}

In Fig.\ref{fig14}, we study the impact of large-scale changes of $k_{1,m}$ and $k_{2,m}$, which represent the memory of the mobility model of terminals. At first, $k_{1,m}=k_{2,m}=0.5$ for $m\in\mathcal{M}$. In the $60,000$th, $80,000$th, $100,000$th, and $120,000$th episode, $k_{1,m}$ and $k_{2,m}$ of all terminals simultaneously change to 1, back to 0.5, change to 0, and back to 0.5 again. For the first three changes, SAC-TR can adapt instantaneously. For the last change, the return first drops drastically and then ascends to a stable value rapidly. Fig.\ref{fig13} and \ref{fig14} demonstrate that SAC-TR has good adaption to unexpected changes of the environment. In reality, even if there exists a certain degree of difference between the model of training samples and the real environment, SAC-TR can also handle it.

\section{Conclusion}\label{conclusion}
In this paper, we study an UAV-assisted wireless powered MEC system for mobile terminals. By combining the computation rate and a fairness index in our objective function, we aim to jointly optimize and continuously control the UAV trajectory and the resource allocation of terminals. An SAC-based algorithm, named SAC-TR, is proposed for trajectory planning and resource allocation to solve this complex high-dimensional DRL task. In SAC-TR, reward is designed as a heterogeneous function including a computation reward and an arrival reward. The computation reward integrates a fairness index to improve the computation rate while guaranteeing fairness, and the arrival reward based on the progress estimate guides the UAV to reach the specified destination and promotes convergence. Simulation results show that SAC-TR can converge stably and adapt to drastic changes of the environment quickly. Compared to widely-used benchmarks, such as the HFH trajectory, the straight trajectory, the greedy algorithm and the random algorithm, the performance of SAC-TR exceeds or approaches them in various situations.

% or


% \appendices

% % you can choose not to have a title for an appendix
% % if you want by leaving the argument blank
% \section{}

% % use section* for acknowledgment
% \section*{Acknowledgment}

\ifCLASSOPTIONcaptionsoff
  \newpage
\fi


\bibliographystyle{IEEEtran}
\bibliography{IEEEabrv,UAV-ref}

% % biography section
% % 
% % If you have an EPS/PDF photo (graphicx package needed) extra braces are
% % needed around the contents of the optional argument to biography to prevent
% % the LaTeX parser from getting confused when it sees the complicated
% % \includegraphics command within an optional argument. (You could create
% % your own custom macro containing the \includegraphics command to make things
% % simpler here.)
% %\begin{IEEEbiography}[{\includegraphics[width=1in,height=1.25in,clip,keepaspectratio]{mshell}}]{Michael Shell}
% % or if you just want to reserve a space for a photo:

% \end{IEEEbiography}


% % insert where needed to balance the two columns on the last page with
% %\newpage


% % You can push biographies down or up by placing
% % a \vfill before or after them. The appropriate
% % use of \vfill depends on what kind of text is
% % on the last page and whether or not the columns
% % are being equalized.

% \vfill

% \enlargethispage{-5in}

\end{document}